\documentclass[twocolumn,english,aps,prl,showpacs,twocolumns]{revtex4-1}
\usepackage[T1]{fontenc}
\usepackage[latin9]{inputenc}
\usepackage{color}
\usepackage{amsmath}
\usepackage{amssymb}
\usepackage{graphicx}
\usepackage{esint}

\makeatletter

\usepackage {url}
\usepackage {lineno}

\makeatother

\usepackage{babel}
\begin{document}

\title{Rydberg-interaction gates via adiabatic passage and phase control
of driving fields}

\author{Huaizhi Wu, Xi-Rong Huang, Chang-Sheng Hu, Zhen-Biao Yang, and Shi-Biao
Zheng }

\address{Fujian Key Laboratory of Quantum Information and Quantum Optics and
Department of Physics, Fuzhou University, Fuzhou 350116, P. R. China}
\begin{abstract}
In this paper, we propose two theoretical schemes for implementation
of quantum phase gates by engineering the phase sensitive dark state
of two atoms subjected to Rydberg-Rydberg interaction. Combining the
conventional adiabatic techniques and newly developed approaches of
phase control, a feasible proposal for implementation of a geometric
phase gate is presented firstly, where the conditional phase shift
(Berry phase) is achieved by adiabatically and cyclically changing
the parameters of the driving fields. Here, we find that the geometric
phase acquired is related to the way how the relative phase is modulated.
In the second scheme, the system Hamiltonian is adiabatically changed
in a noncyclic manner, so that the acquired conditional phase is not
a Berry phase. A detailed analysis of the experimental feasibility
and the effect of decoherence is also given. The proposed schemes
provide new perspectives for adiabatic manipulation of interacting
Rydberg systems with tailored phase modulation.
\end{abstract}
\maketitle

\section{Introduction}

Neutral atoms in highly excited and long-lived Rydberg states are
considered as the ideal architecture for quantum information processing
since it provides strongly interatomic interaction on demand, while
keeps interacting with the environment weakly \cite{Saffman2010,Browaeys2016}.
There have been numerous proposals to use Rydberg-Rydberg interactions
for implementation of quantum logic gates \cite{Cozzini2006375,Brion2007,Muller2009,PhysRevA.82.034307,Rao2014,Keating2015,Beterov2016a},
quantum error correction \cite{Zeiher2015,PhysRevLett.100.110506},
quantum algorithms \cite{Chen2011,Sanders2014,Petrosyan} and quantum
repeater \cite{Han2010,PhysRevA.81.052329,PhysRevA.85.042324,Solmeyer2015a}.
By following the pioneering works proposed by Jaksch et al. \cite{PhysRevLett.85.2208}
and Lukin et al. \cite{PhysRevLett.87.037901}, promising schemes
for realizing two-qubit controlled-Z and controlled-NOT gates that
rely on dynamical control of dipolar coupling and intrinsic Förster
interaction have been widely studied in both the Rydberg blockade
\cite{Muller2009,Isenhower_PRL2010,Zhang_PRA12_FideRydGate,Muller2014,Maller2015}
and antiblockade regimes \cite{Su2017,Petrosyan2014}. Therein, the
validity of the gate operations is predominantly determined by the
detailed laser parameters as well as the Rydberg interaction strength.
Experimental demonstrations in producing quantum entanglement of few
Rydberg atoms \cite{Barredo2014,Zeng2017} and two-qubit logic operations
\cite{Maller2015} have recently made great progress by addressing
the system's evolutional dynamics, however, the fidelity achieved
to date is significantly limited by the imprecise control of experimental
parameters. 

The requirement of precise control of coherent dynamics can be avoided
by using the adiabatic techniques, such as stimulated Raman adiabatic
passage (STIRAP) and adiabatic rapid passage (ARP), where the sensitivity
to imprecise Rabi control and other experimental perturbations is
strongly suppressed \cite{Vitanov2017}. The theoretical proposals
based on the STIRAP and the ARP have been proposed for coherent population
transfer \cite{PhysRevA.84.023413,Yan2011,Qian2015a,Tian2015,Petrosyan2015d},
preparation of entangled states \cite{Moller2008,Petrosyan}, and
implementation of quantum logic gates \cite{Goerz2014,PhysRevA.88.010303,Rao2014,Keating2015}
with Rydberg atoms, which exhibit robustness properties against moderate
fluctuations of experimental parameters. Furthermore, the adiabatic
technique alternatively provides a chance for geometric manipulation
of Rydberg systems \cite{Moller2008,PhysRevA.86.032323,PhysRevA.88.010303},
which is naturally robust against certain control errors \cite{Zheng2015,Zheng2016}
and is a promising approach for implementation of a built-in fault-tolerant
two-qubit logic gate.

Here, we put forward two new schemes for implementing quantum phase
gates via adiabatic passage and phase control of the driving fields.
The first scheme is based on the geometric manipulation of the system's
Hamiltonian in the parameter space. In contrast to the previously similar
approach \cite{Moller2008}, the geometric phase acquired here is
not due to the variance of the phase difference of the control pulses,
and is alternatively accumulated by changing the phases of the driving
fields in step and keeping the phase difference null. Remarkably,
we find that the geometric phase acquired is strongly dependent on
the way how the relative phase is modulated. In the second scheme,
neither is the conditional phase shift of dynamical origin since the
qubit system evolves in the dark state space, nor is it a Berry adiabatic
phase as the system Hamiltonian is not cyclically changed. The conditional
phase arises from the adiabatic manipulation of the dark state with
staircase phase control. The experimental feasibility, gate fidelity
and docoherence effect for the proposed schemes are carefully studied.

This paper is organized as follows. In Sec. II, we propose the level
addressing scheme for two neutral atoms interacting via
the Rydberg-Rydberg interaction and examine the role of the phases of driving fields
in adiabatic control. In Sec. III, the schemes for implementing conditional
phase gates based on Berry phase and non-Berry adiabatic phase are
presented. In Sec. IV, we provide a detailed discussion about the
experimental feasibility of the two schemes. In Sec. V, the effect
of atomic spontaneous emission and interatomic force on the gate fidelity
is studied. The conclusion appears in Sec. VI.

\section{Dark state of two interacting Rydberg atoms}

\begin{figure}
\begin{centering}
\includegraphics[width=1\columnwidth]{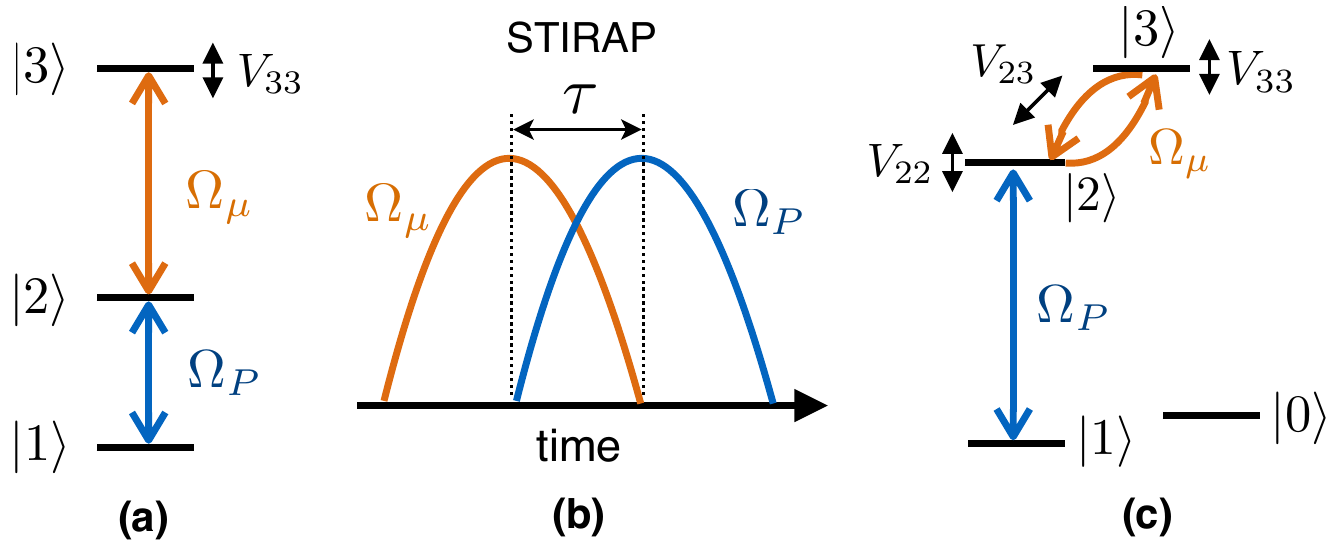}
\par\end{centering}
\caption{\label{fig:level structure}(color online) (a) Schematic level configuration.
The Rydberg state $|3\rangle$ is excited from the ground state $|1\rangle$
via an intermediate state $|2\rangle$ with two lasers of optical
frequencies. Double excitation of the Rydberg state $|3\rangle$ will
be shifted by $V_{33}$ due to the interatomic interaction. $\Omega_{p}$
and $\Omega_{\mu}$ are Rabi frequencies for the transitions $|1\rangle\leftrightarrow|2\rangle$
and $|2\rangle\leftrightarrow|3\rangle$, respectively. (b) STIRAP
pulse sequence applied to the interacting Rydberg atoms with $\tau$
being the overlapping time. (c) Addressing scheme with multiple Rydberg
levels. The ground state $|1\rangle$ is resonantly coupled to the
Rydberg state $|2\rangle$ via single photon transition with Rabi
frequency $\Omega_{p}$ and the atomic transition between Rydberg
states $|2\rangle$ and $|3\rangle$ is driven by a microwave field
with Rabi frequency $\Omega_{\mu}$. The ground state $|0\rangle$
is the auxiliary qubit state. $V_{22}$ , $V_{23}$ , and $V_{33}$
are energy shifts induced by the Rydberg-Rydberg interaction. }
\end{figure}
\begin{figure}
\begin{centering}
\includegraphics[width=1\columnwidth]{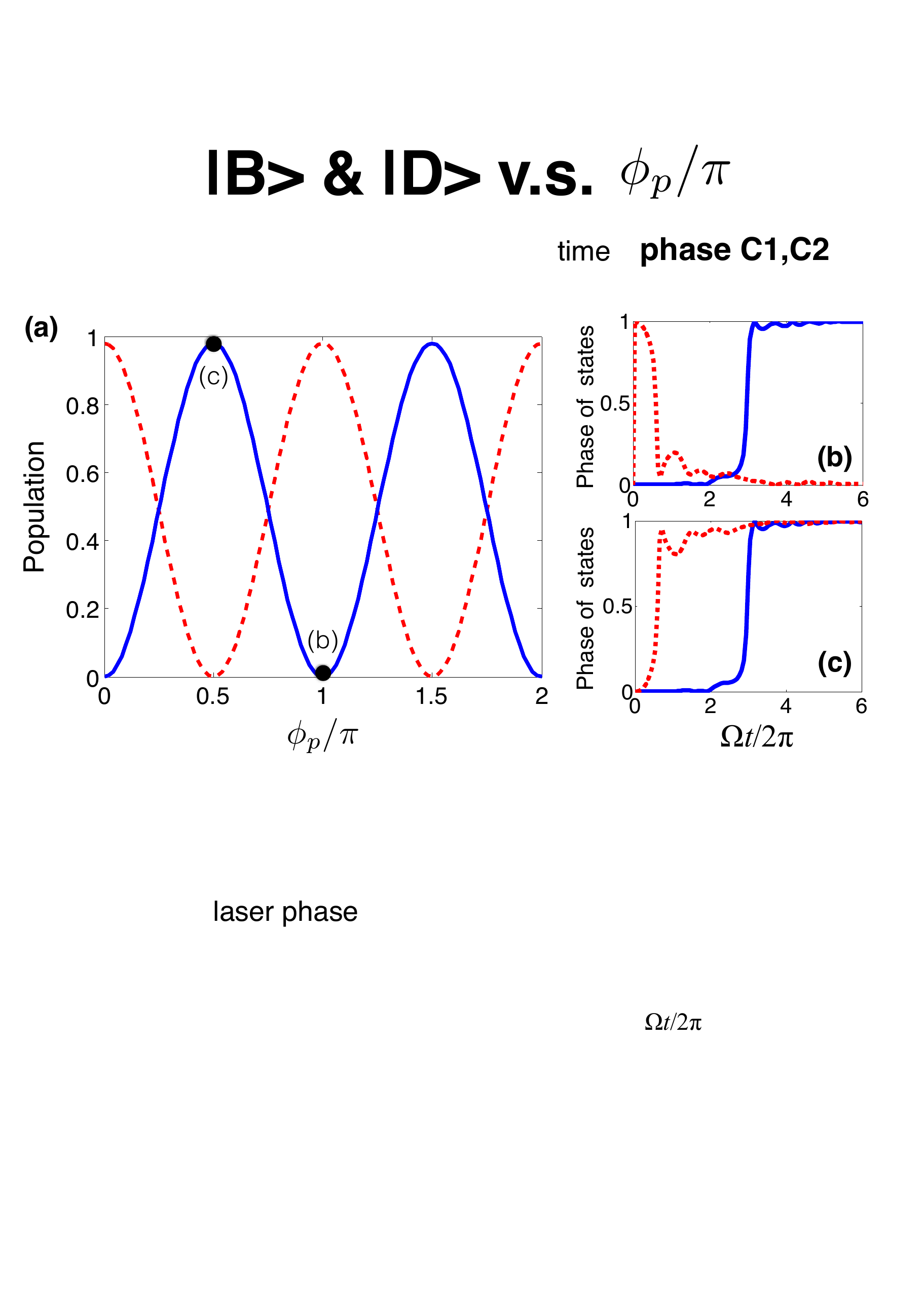}
\par\end{centering}
\caption{\label{fig: wavelike}(color online) (a) $\phi_{p}$-dependent wavelike
interference fringes in the probabilities of $|EPR\rangle_{s}$ (solid
blue) and $|EPR\rangle_{as}$ (dashed red). (b), (c) Numerically calculated
time-dependencies of the phases (divided by $\pi$) of the states
$|\phi_{1}\rangle$ (solid blue) and $|\phi_{4}\rangle$ (dashed red)
for $\phi_{p}=0$ and $\phi_{p}=\pi/2$, respectively. The two-atom
system is initially in the state $|\phi_{1}\rangle$ and adiabatically
evolves along $|d_{2}(t)\rangle$. The Rabi frequencies are modeled
by sine-function pulses \textcolor{black}{$\Omega_{p}(t)=\Omega\textrm{sin}(\frac{\pi}{2\tau}t)e^{i\phi_{p}},$
$\Omega_{\mu}(t)=\Omega|\textrm{cos}(\frac{\pi}{2\tau}t)|$ with $0\leqslant t\leqslant\tau$}.
We fix units of \textcolor{black}{$\Omega=1$, and set $\Omega\tau/2\pi=6$,
$V_{23}/\Omega=1.1$, and $V_{33}/\Omega=0.9$}.}
\end{figure}

We first introduce the schematic description of the system. Consider
a pair of identical three-level atoms with a ground state $|1\rangle$,
an intermediate state $|2\rangle$, and a highly excited Rydberg state
$|3\rangle$, see Fig. \ref{fig:level structure}(a), which are trapped
in optical tweezers or optical lattices. Two excitation lasers of
optical frequencies resonantly drive the atomic transitions $|1\rangle\leftrightarrow|2\rangle$
and $|2\rangle\leftrightarrow|3\rangle$ with the Rabi frequencies
$\Omega_{p}\equiv|\Omega_{p}|e^{i\phi_{p}}$ and $\Omega_{\mu}\equiv|\Omega_{\mu}|e^{-i\phi_{\mu}}$
(taken as complex number), respectively. The atoms experience an energy
shift $V_{33}$ when both atoms are excited to the Rydberg state $|3\rangle$.
The total Hamiltonian of the system in the rotating wave approximation
is
\begin{equation}
\begin{aligned}\mathcal{H}_{R}= & \mathcal{H}_{1}\otimes\mathcal{I}_{2}+\mathcal{I}_{1}\otimes\mathcal{H}_{2}+V_{33}|3\rangle_{1}|3\rangle_{22}\langle3|_{1}\langle3|,\end{aligned}
\end{equation}
with (from now on, we put $\hbar=1$)
\begin{equation}
\mathcal{H}_{i}=\Omega_{p}|2\rangle_{ii}\langle1|+\Omega_{\mu}|3\rangle_{ii}\langle2|+h.c.,\;i=1,2.
\end{equation}
In terms of the symmetric two-atomic basis states spanned by $\{|\phi_{j}\rangle\}$,
$j=1,...,6$, with
\begin{eqnarray*}
|\phi_{1}\rangle & = & |1\rangle_{1}|1\rangle_{2},\\
|\phi_{2}\rangle & = & \textrm{ }\frac{1}{\sqrt{2}}(|1\rangle_{1}|2\rangle_{2}+|2\rangle_{1}|1\rangle_{2}),\\
|\phi_{3}\rangle & = & \frac{1}{\sqrt{2}}(|1\rangle_{1}|3\rangle_{2}+|3\rangle_{1}|1\rangle_{2}),\\
|\phi_{4}\rangle & = & \textrm{ }|2\rangle_{1}|2\rangle_{2},\\
|\phi_{5}\rangle & = & \frac{1}{\sqrt{2}}(|2\rangle_{1}|3\rangle_{2}+|3\rangle_{1}|2\rangle_{2}),\\
|\phi_{6}\rangle & = & \textrm{ }|3\rangle_{1}|3\rangle_{2},
\end{eqnarray*}
$\mathcal{H}_{R}$ can be rewritten as
\begin{equation}
\mathcal{H}_{R}=\left[\begin{array}{cccccc}
0 & \sqrt{2}\Omega_{p}^{*} & 0 & 0 & 0 & 0\\
\sqrt{2}\Omega_{p} & 0 & \Omega_{\mu}^{*} & \sqrt{2}\Omega_{p}^{*} & 0 & 0\\
0 & \Omega_{\mu} & 0 & 0 & \Omega_{p}^{*} & 0\\
0 & \sqrt{2}\Omega_{p} & 0 & 0 & \sqrt{2}\Omega_{\mu}^{*} & 0\\
0 & 0 & \Omega_{p} & \sqrt{2}\Omega_{\mu} & 0 & \sqrt{2}\Omega_{\mu}^{*}\\
0 & 0 & 0 & 0 & \sqrt{2}\Omega_{\mu} & V_{33}
\end{array}\right].\label{eq:HR}
\end{equation}
There exists a nondegenerate eigenspace and a unique dark state for
the Hamiltonian $\mathcal{H}_{R}$, which is given by
\begin{eqnarray}
|d_{2}(t)\rangle & \propto & (|\Omega_{\mu}|^{2}-|\Omega_{p}|^{2})|\phi_{1}\rangle+\Omega_{p}^{2}|\phi_{4}\rangle-\sqrt{2}\Omega_{\mu}\Omega_{p}|\phi_{3}\rangle.\nonumber \\
\label{eq:ds2_unm}
\end{eqnarray}
Expressing the relative strength of the two Rabi frequencies $\Omega_{p}$,
$\Omega_{\mu}$ as $\textrm{tan}\theta=|\Omega_{p}|/|\Omega_{\mu}|$
and keeping their phases nonvanishing, the Eq.(\ref{eq:ds2_unm})
after normalization is rewritten as follows:
\begin{eqnarray}
|d_{2}(t)\rangle & = & \mathcal{N}^{-1}[(\textrm{cos}^{2}\theta-\textrm{sin}^{2}\theta)|\phi_{1}\rangle+\textrm{sin}^{2}\theta e^{i2\phi_{p}}|\phi_{4}\rangle\nonumber \\
 &  & -\sqrt{2}\textrm{sin}\theta\textrm{cos}\theta e^{-i(\phi_{\mu}-\phi_{p})}|\phi_{3}\rangle],\label{eq:ds2_norm}
\end{eqnarray}
where
\[
\textrm{cos}\theta=\frac{|\Omega_{\mu}|}{\sqrt{|\Omega_{\mu}|^{2}+|\Omega_{p}|^{2}}},
\]
\[
\textrm{sin}\theta=\frac{|\Omega_{p}|}{\sqrt{|\Omega_{\mu}|^{2}+|\Omega_{p}|^{2}}},
\]
and
\[
\mathcal{N}=\sqrt{\textrm{cos}^{4}\theta+2\textrm{sin}^{4}\theta}.
\]
\textcolor{black}{The }Eq.(\ref{eq:ds2_norm}) has a similar form
to the dark state firstly studied by Møller et al. \textcolor{black}{\cite{Moller2008},
where a time-dependent relative phase $\phi_{r}(t)\equiv\phi_{\mu}-\phi_{p}$
is found to be relevant for acquisition of geometric phases, except
that the Eq.} (\ref{eq:ds2_norm})\textcolor{black}{{} contains an additional
exponential factor $e^{i2\phi_{p}}$ for $|\phi_{4}\rangle$. It implies
that by setting $\phi_{p}=0$ or $\phi_{p}=\pi/2$, the two-atom system
will transfer to the anti-symmetric superposition state} $|EPR\rangle_{as}=(|\phi_{1}\rangle-|\phi_{4}\rangle)/\sqrt{2}$
\textcolor{black}{or the symmetric superposition state $|EPR\rangle_{s}=(|\phi_{1}\rangle+|\phi_{4}\rangle)/\sqrt{2}$
by adiabatically following the dark state with $\theta$ changing
from 0 to $\pi/2$ (see later discussion for detail). This is numerically
confirmed by examining the probability for detecting the two-atomic
states $|EPR\rangle_{s,as}$ after the applied STIRAP pulse sequence
{[}see Fig.\ref{fig:level structure}(b){]}, which is a sinusoidal
function of $\phi_{p}$ exhibiting wavelike interference fringes,
as shown in Fig. \ref{fig: wavelike}. }The co-existence of the phase
factors $1,$\textcolor{black}{{} $e^{-i\phi_{r}}$ and $e^{i2\phi_{p}}$
in the superposition coefficients for the three components $|\phi_{1}\rangle$,
$|\phi_{3}\rangle$ and $|\phi_{4}\rangle$, respectively, can significantly
modify the }geometric phases acquired during adiabatic evolution assisted
by phase control and \textcolor{black}{can find special use for construction
of quantum logic gates.}

\textcolor{black}{The level configuration {[}as in Fig. }\ref{fig:level structure}\textcolor{black}{(a){]}
including a single Rydberg state suffers from }an irreversible spontaneous
decay\textcolor{black}{{} since the optically excited intermediate state}
$|2\rangle$ has a short lifetime, therefore adiabatic manipulation
of the (unstable) dark state becomes not experimentally feasible (see
section V for further discussion). To avoid the defect, we then consider
atoms with two ground hyperfine states $|0\rangle$ and $|1\rangle$
and two Rydberg states $|2\rangle$ and $|3\rangle$, see Fig. \ref{fig:level structure}(c).
The atomic transition $|1\rangle\leftrightarrow|2\rangle$ is resonantly
excited by a single-photon field $\Omega_{p}$ and the transition
$|2\rangle\leftrightarrow|3\rangle$ is driven by a microwave field
$\Omega_{\mu}$. The auxiliary level $|0\rangle$ is introduced as
a qubit information for the later discussed gate protocols. While
both atoms are excited to the Rydberg states, two relevant interparticle
interactions are involved, i.e. the van der Waals (vdW) interaction
$V_{22}$ ($V_{33}$) between the states $|2\rangle$ ($|3\rangle$)
and the exchange dipole-dipole interaction (DDI) $V_{23}$ between
an atom in $|2\rangle$ and another in $|3\rangle$. When including
the Rydberg-Rydberg interaction, the two-atom Hamiltonian governing
the temporal evolution of the compound system takes the form 

\begin{equation}
\begin{aligned}\mathcal{H}'_{R}= & \mathcal{H}_{1}\otimes\mathcal{I}_{2}+\mathcal{I}_{1}\otimes\mathcal{H}_{2}+V_{33}|3\rangle_{1}|3\rangle_{22}\langle3|_{1}\langle3|\\
 & +V_{22}|2\rangle_{1}|2\rangle_{22}\langle2|_{1}\langle2|+V_{23}(|2\rangle_{1}|3\rangle_{22}\langle3|_{1}\langle2|\\
 & +|3\rangle_{1}|2\rangle_{22}\langle2|_{1}\langle3|).
\end{aligned}
\label{eq:Hamil_2RydS}
\end{equation}

If the two atoms only weakly interact with each other while they are
in the state $|2\rangle$ (e.g., due to a dispersive Förster process)
such that the vdW shift $V_{22}$ becomes negligible comparing with
other Rydberg interaction energies $V_{33}$, $V_{23}$, i.e. $V_{22}\ll V_{23,}V_{33}$,
then  the Hamiltonian  $\mathcal{H}'_{R}$ with $V_{22}\rightarrow0$
has one dark state, which is exactly given by Eq. (\ref{eq:ds2_norm}).
The interatomic DDI $V_{23}$ does not shift the zero eigenenergy
and change the form of the dark state. Therefore, the single-Rydberg-level effects with respect to $|d_{2}\rangle$ hold true for the multiple-Rydberg-level model as long as the adiabatic condition is well guaranteed, and adiabatic control of the dark state becomes more feasible for long radiative lifetime of the Rydberg levels.

In another parameter regime where the interaction between the Rydberg
states $|3\rangle$ is sufficiently weak comparing with the vdW shift
$V_{22}$ and the DDI strength $V_{23}$, i.e. $V_{33}\ll V_{22,}V_{23},$
by setting $V_{33}=0$ we again find a dark state for $\mathcal{H}'_{R}$,
but with a different form
\begin{eqnarray}
|d'_{2}(t)\rangle & = & \textrm{cos}^{2}\theta e^{i2\phi_{r}}|\phi_{1}\rangle+\textrm{sin}^{2}\theta|\phi_{6}\rangle\nonumber \\
 &  & -\sqrt{2}\textrm{sin}\theta\textrm{cos}\theta e^{i\phi_{r}}|\phi_{3}\rangle,\label{eq:ds2_v33}
\end{eqnarray}
which can be exactly expressed as the direct product of the dark states
for the single-atom Hamiltonian $\mathcal{H}_{i}$ ($i=1,2$), i.e.
$|d'_{2}(t)\rangle=|d_{1}(t)\rangle_{1}\otimes|d_{1}(t)\rangle_{2}$,
with

\begin{equation}
|d_{1}(t)\rangle_{i}=\textrm{cos}\theta e^{i\phi_{r}}|1\rangle_{i}-\textrm{sin}\theta|3\rangle_{i}.\label{eq:ds1}
\end{equation}
In this case, the relative phase $\phi_{r}$ is the only degree of
freedom for phase modulation during the system's adiabatic evolution
along $|d'_{2}(t)\rangle$. 

If we further assume that $V_{22}=V_{33}=0$ but with $V_{23}\neq0$,
the zero-energy eigenstate for the two-atom Hamiltonian $\mathcal{H}'_{R}$
can then be written as the superposition of the degenerated dark states
$|d_{2}(t)\rangle$ and $|d'_{2}(t)\rangle$. A finite Rydberg interaction
strength $V_{22}$ or $V_{33}$ between the states $|2\rangle$ or
$|3\rangle$ results in the removal of the degeneracy, which cannot
occur with only the DDI due to the missing component $|\phi_{5}\rangle$.

Suppose that the Hamiltonian $\mathcal{H}'_{R}(t)$ is time dependent
through the set of parameters $\mathbf{R}(t)=(\theta(t),\phi_{p}(t),\phi_{r}(t))$
and the interacting two-atom system is initially in the ground eigenstate
$|g(\mathbf{R}(0))\rangle$ of the instantaneous $\mathcal{H}'_{R}(t=0)$.
If $\mathbf{R}(t)=(\theta(t),\phi_{p}(t),\phi_{r}(t))$ is modulated
under the condition
\[
|\langle e(t)|\frac{d\mathcal{H}'_{R}}{dt}|g(t)\rangle|\ll|E_{e}-E_{g}|^{2}
\]
 such that the Hamiltonian is adiabatically changed along a closed
curve $\boldsymbol{C}$ in the parameter space (i.e. $\mathbf{R}(T)=\mathbf{R}(0)$),
where $|e\rangle$ is any one of the instantaneous excited state,
then the system will keep in the ground state and acquire a purely
geometric phase $\text{\ensuremath{\varphi}}_{g}$ in additional to
the usual dynamical phase $\text{\ensuremath{\varphi}}_{d}$:
\begin{equation}
|g(\mathbf{R}(T))\rangle=exp\{i[\varphi_{g}(T)+\varphi_{d}(T)]\}|g(\mathbf{R}(0))\rangle\text{,}\label{eq:adiabatic_evolution}
\end{equation}
where 
\begin{equation}
\varphi_{g}=i\oint_{\boldsymbol{C}}d\mathbf{R}\cdot\langle g(\mathbf{R}(t))|\nabla_{\mathbf{R}}|g(\mathbf{R}(t))\rangle\label{eq:geo_phase_def}
\end{equation}
and
\begin{equation}
\varphi_{d}(T)=-\int_{0}^{T}E_{d}(\mathbf{R}(t))dt,\label{eq:dyn_phase_def}
\end{equation}
which is vanished for a dark state $|g(t)\rangle=|d(\mathbf{R}(t))\rangle$
with zero eigenenergy $E_{g}=0$.

\section{Schemes for implementing Controlled-Z gates via adiabatic passage}

We encode qubit information on the ground state $|1\rangle$ and the
auxiliary level $|0\rangle$ that is uncoupled from any pulse sequences
of the control field. Thus, the computational basis states are given
by $\{|0\rangle_{1}|0\rangle_{2},\textrm{ |0\ensuremath{\rangle_{1}}|1\ensuremath{\rangle_{2}},}\textrm{ }|1\rangle_{1}|0\rangle_{2},\textrm{ |1\ensuremath{\rangle_{1}}|1\ensuremath{\rangle_{2}}}\}$.
The controlled-Z gate is implemented by applying a counterintuitive
pulse sequence and by modulating the phases of the control fields. 

\begin{figure}
\begin{centering}
\includegraphics[width=0.9\columnwidth]{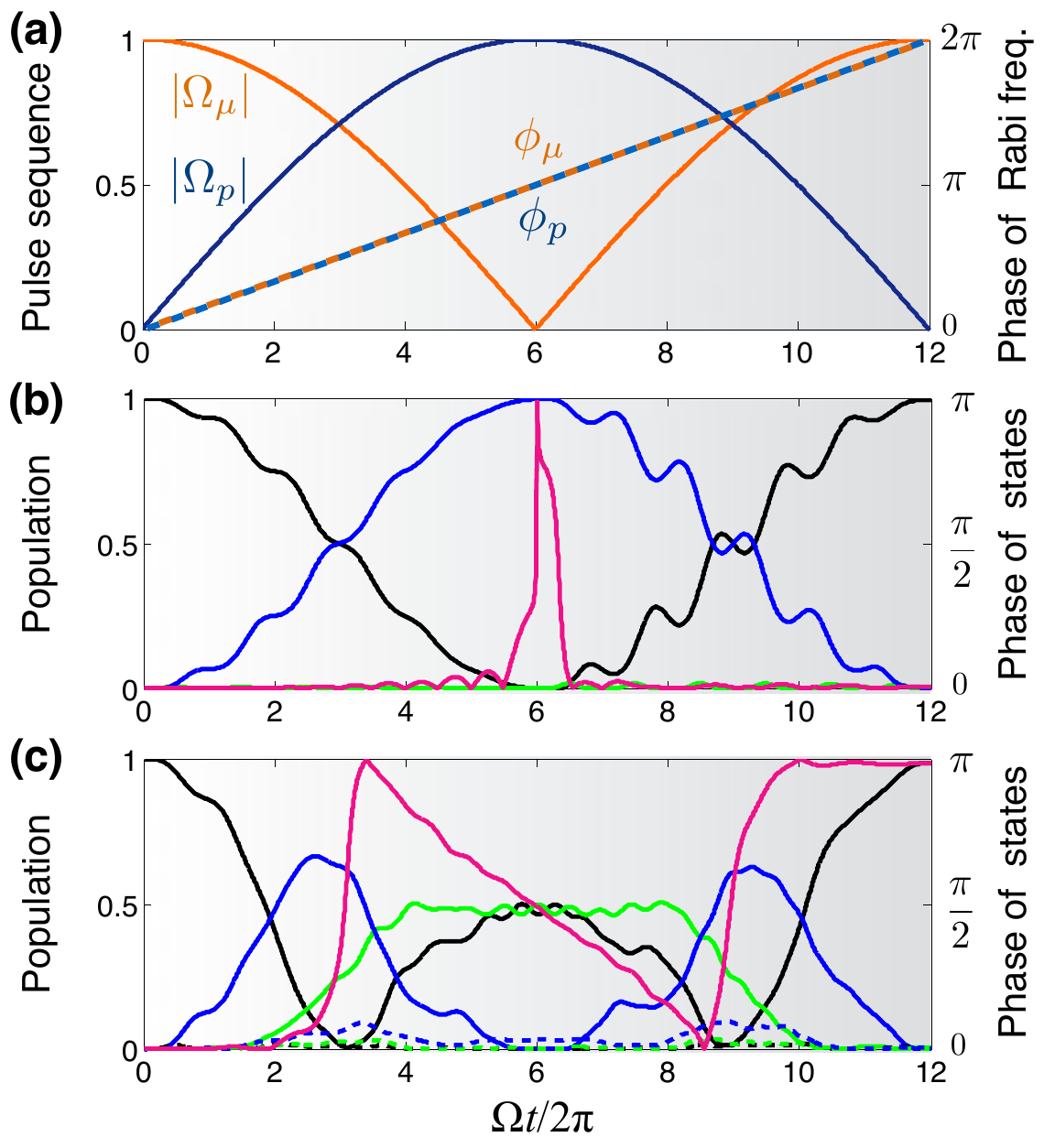}
\par\end{centering}
\caption{\label{fig:population_geo}(color online) (a) The amplitudes $|\text{\ensuremath{\Omega}}_{p}(t)|$,
$|\Omega_{\mu}(t)|$ and phases $\phi_{p}$, $\phi_{\mu}$ of Rabi
frequencies as a function of rescaled time. (b) Time dependent population
of the states $|0\rangle_{1}|1\rangle_{2}$ ($|1\rangle_{1}|0\rangle_{2}$)
(black), $|0\rangle_{1}|3\rangle_{2}$ ($|3\rangle_{1}|0\rangle_{2}$)
(blue) and $|0\rangle_{1}|2\rangle_{2}$ ($|2\rangle|0\rangle_{2}$)
(green), and the phase of state $|0\rangle_{1}|1\rangle_{2}$ ($|1\rangle_{1}|0\rangle_{2}$)
(magenta) for the system initially in $|d_{1}(0)\rangle$. (c) Time dependent
population of the states $|1\rangle_{1}|1\rangle_{2}$ (solid black),
$|\phi_{2}\rangle$ (dash black), $|\phi_{3}\rangle$ (solid blue),
$|\phi_{4}\rangle$ (solid green), $|\phi_{5}\rangle$ (dash green)
and $|\phi_{6}\rangle$ (dash blue), and the phase of state $|1\rangle_{1}|1\rangle_{2}$
(magenta) for the system initially in $|d_{2}(0)\rangle$. We fix units of
\textcolor{black}{$\Omega=1$, and set $\Omega\tau/2\pi=6$, $V_{23}/\Omega=1.1$,
and $V_{33}/\Omega=0.9$}.}
\end{figure}

\textbf{Scheme 1.} Geometric phase gate with the phases of the Rabi
frequencies varying in step (i.e. $\phi_{r}=$const.). It has been
realized that the dark states (\ref{eq:ds2_norm}) and (\ref{eq:ds1})
under adiabatic evolution can acquire the geometric phases $\varphi_{2}=2\int\textrm{sin}^{2}\theta\textrm{cos}^{2}\theta(\textrm{cos}^{4}\theta+2\textrm{sin}^{4}\theta)^{-1}d\phi_{r}$
and $\varphi_{1}=\int\textrm{sin}^{2}\theta d\phi_{r}$, respectively,
for a nonvanishing and time-dependent relative phase $\phi_{r}(t)$
\cite{Moller2008}. In contrast, we find that the two-atom dark state
(\ref{eq:ds2_norm}) can acquire a Berry phase even though the relative
phase is kept invariant.

Suppose the two-atom system is initially in $|d_{2}(0)\rangle=|1\rangle_{1}|1\rangle_{2}$
(i.e. $\textrm{cos}\theta=1$) and the phases of the driving fields
are $\phi_{p}(0)=0$, $\phi_{\mu}(0)=0$ without loss of generality.
The time-dependent amplitudes of the Rabi frequencies are chosen as
($0\leq t\leq2\tau$)
\begin{equation}
|\Omega_{p}(t)|=\Omega\textrm{sin}(\frac{\pi}{2\tau}t),\:|\Omega_{\mu}(t)|=\Omega|\textrm{cos}(\frac{\pi}{2\tau}t)|,\label{eq:pulse_s1}
\end{equation}
which corresponds to $\theta(t)$ varying from $0$ to $\pi/2$ and
the corresponding reverse process. The phases $\phi_{p,\mu}(t)$ are
synchronized with each other in real time and have a simply linear
time dependence $\phi_{p,\mu}(t)=\text{\ensuremath{\pi}}t/\tau$.
Therefore, the system makes a cyclic evolution with starting point
and ending point $\theta=0$, see temporal evolution of the probability
amplitudes and the phases of the relevant states as shown in Fig.
\ref{fig:population_geo}. The geometric phase $\varphi'_{2}$ (i.e.
the Berry phase) accumulated during the adiabatic process can be calculated
by using the standard formula Eq. (\ref{eq:geo_phase_def}). Since
$\phi_{r}$ remains zero at any time, the relevant parameter space
reduces to $\mathbf{R}(t)=(\theta(t),\phi_{p}(t))$. Thus, we have
\begin{eqnarray}
\varphi'_{2} & = & -\oint_{C}\frac{2\textrm{sin}^{4}\theta}{\textrm{cos}^{4}\theta+2\textrm{sin}^{4}\theta}d\phi_{p}\nonumber \\
 & = & -\oint_{C}\frac{4\textrm{sin}^{4}\theta}{\textrm{cos}^{4}\theta+2\textrm{sin}^{4}\theta}d\text{\ensuremath{\theta}}\label{eq:geo_phase_s1}
\end{eqnarray}
for $d\phi_{p}(t)/d\theta(t)=2$ taken in our example. Apart from
that, while the system is initially in the state $|0\rangle_{1}|0\rangle_{2}$,
$|0\rangle_{1}|1\rangle_{2}$ or $|1\rangle_{1}|0\rangle_{2}$, no
geometric phases can be acquired during the cyclic evolution. The
sudden increase of the phases of $|0\rangle_{1}|1\rangle_{2}$ ($|1\rangle_{1}|0\rangle_{2}$)
around $t=\tau$ is due to imperfect state transfer and is automatically
eliminated at the end of the pulse sequence. Thus, we have successfully
implemented a controlled phase gate based on the conditionally geometric
phase shift:
\begin{gather}
|0\rangle_{1}|0\rangle_{2}\longrightarrow|0\rangle_{1}|0\rangle_{2},\quad|0\rangle_{1}|1\rangle_{2}\longrightarrow|0\rangle_{1}|1\rangle_{2},\nonumber \\
|1\rangle_{1}|0\rangle_{2}\longrightarrow|1\rangle_{1}|0\rangle_{2},\quad|1\rangle_{1}|1\rangle_{2}\longrightarrow e^{i\varphi'_{2}}|1\rangle_{1}|1\rangle_{2}.\label{eq:geo-gate}
\end{gather}
 \textcolor{black}{Since we have guaranteed $\phi_{r}=const.$ during
the adiabatic evolution, the required solid angle for obtaining the
geometric phase $\varphi'_{2}$ is induced by the} concurrency control
of\textcolor{black}{{} $\phi_{p}$ and $\phi_{\mu}$, where an additional
reference oscillator should be included. }While for the system initially
being in the two-atom dark state $|d_2'(t)\rangle$ under the condition
$V_{33}\ll V_{22,}V_{23}$, no geometric phases can be acquired since
$\phi_{r}$ is invariant.

\begin{figure}
\begin{centering}
\includegraphics[width=0.85\columnwidth]{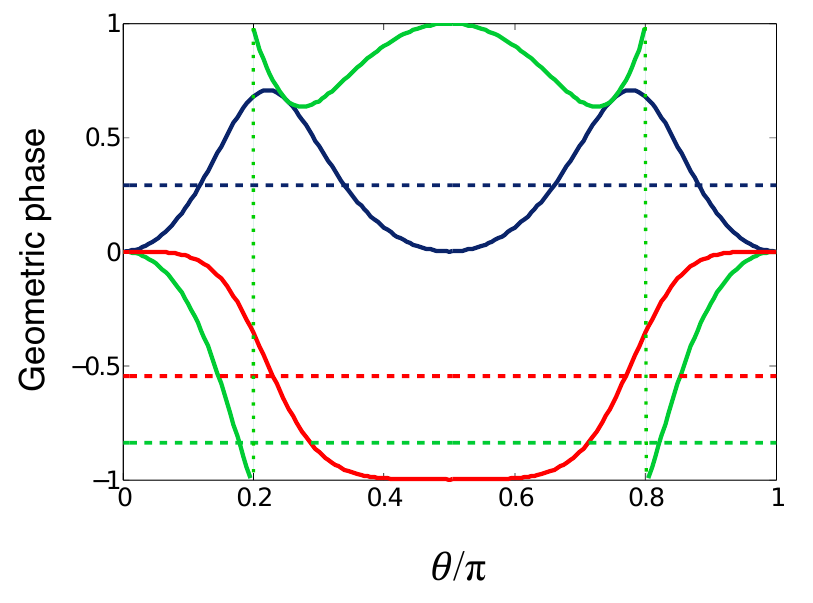}
\par\end{centering}
\caption{\label{fig:geo_vs_theta}(color online) The acquired geometric phases
$\varphi_{2}$, $\varphi_{2}'$ and $\varphi_{2}''$ versus $\theta$.
The system in the initial state $|d_{2}(0)\rangle=|1\rangle_{1}|1\rangle_{2}$
is adiabatically taken to the superposition state with a given $\theta$
followed by sweeping the phase of the controlled fields: $\phi_{p}$,
$\phi_{\mu}$: $0\rightarrow\pi$ (red), $\phi_{p}=0$, $\phi_{\mu}$:
$0\rightarrow\pi$ (blue) and $\phi_{p}$: $0\rightarrow\pi$, $\phi_{\mu}=0$
(green), respectively. The amplitudes of the applied pulse sequence
and other parameters are as in Fig. \ref{fig:population_geo}. }
\end{figure}

\textcolor{black}{However, if $\phi_{r}$ becomes time variant, note
that the phases of the Rabi frequencies may be modulated in two fashions,
leading to differently geometric phase shift for the state $|d_{2}\text{\ensuremath{\rangle}}$.
First, $\phi_{p}=const.$ and $\phi_{r}(t)=\phi_{\mu}(t)-\phi_{p}$
is time dependent via $\phi_{\mu}(t)$. In this case, the geometric
phase acquired is exactly given by $\varphi_{2}$. While for the other
case where $\phi_{\mu}=const.$ and $\phi_{r}(t)=\phi_{\mu}-\phi_{p}(t)$
is determined by $\phi_{p}(t)$, the geometric phase acquired is then
alternatively given by
\begin{eqnarray}
\varphi_{2}'' & = & -\oint_{C}\frac{2\textrm{sin}^{2}\theta}{\textrm{cos}^{4}\theta+2\textrm{sin}^{4}\theta}d\phi_{p}\nonumber \\
 & = & -\oint_{C}\frac{4\textrm{sin}^{2}\theta}{\textrm{cos}^{4}\theta+2\textrm{sin}^{4}\theta}d\theta.\label{eq:geo_phase_phip}
\end{eqnarray}
}Thus, we find three different ways of phase control for geometrically
manipulating the interacting two-atom system. A comparison of the
acquired geometric phases for the three cases is shown in Fig. \ref{fig:geo_vs_theta},
from which one can easily read out $\varphi_{2}$, $\varphi_{2}'$
and $\varphi_{2}''$ by $4\theta_{m}\overline{f(\theta)}$, with $\theta_{m}$
and $\overline{f(\theta)}$ being the given $\theta$ with respect
to the preset dark state $d_{2}(\theta_{m},\phi_{p},\phi_{\mu})$ and the
$\theta$ average of the curves on the plots (indicated by dash lines for scheme 1),
respectively.

\begin{figure}
\begin{centering}
\includegraphics[width=0.9\columnwidth]{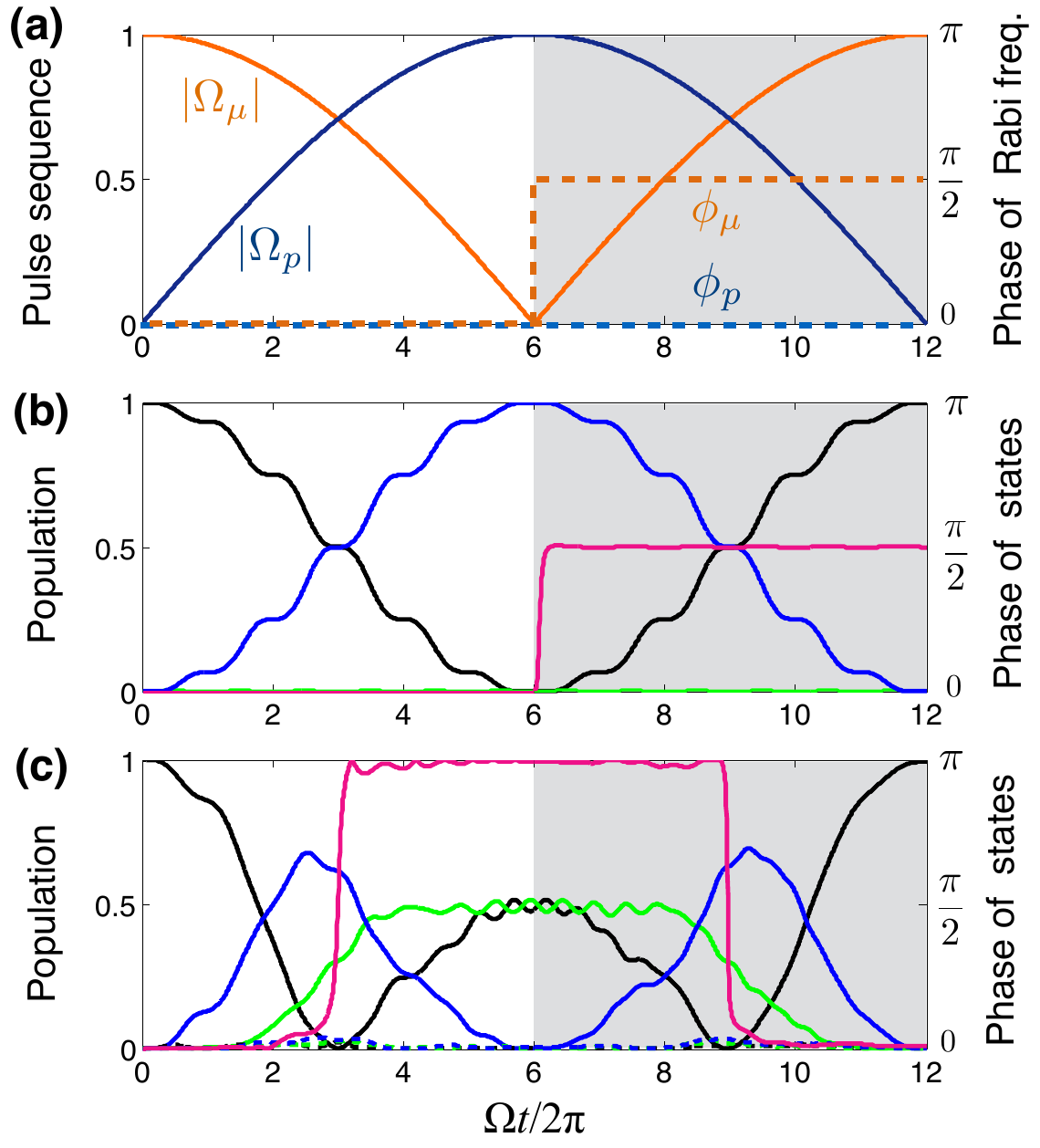}
\par\end{centering}
\caption{\label{fig:population_ng}(color online) (a) The amplitudes $|\text{\ensuremath{\Omega}}_{p}(t)|$,
$|\Omega_{\mu}(t)|$ and phases $\phi_{p}$, $\phi_{\mu}$ of Rabi
frequencies as a function of rescaled time. Without loss of generality,
we set $\phi_{p,\mu}(0)=0$. (b) Time dependence of the population of
states $|0\rangle_{1}|1\rangle_{2}$ ($|1\rangle_{1}|0\rangle_{2}$)
(black), $|0\rangle_{1}|3\rangle_{2}$ ($|3\rangle_{1}|0\rangle_{2}$)
(blue) and $|0\rangle_{1}|2\rangle_{2}$ ($|2\rangle|0\rangle_{2}$)
(green), and the phase of state $|0\rangle_{1}|1\rangle_{2}$ ($|1\rangle_{1}|0\rangle_{2}$)
(magenta) for the system initially in $|d_{1}(0)\rangle$. (c) Time dependence
of the population of states $|1\rangle_{1}|1\rangle_{2}$ (solid black),
$|\phi_{2}\rangle$ (dash black), $|\phi_{3}\rangle$ (solid blue),
$|\phi_{4}\rangle$ (solid green), $|\phi_{5}\rangle$ (dash green)
and $|\phi_{6}\rangle$ (dash blue),
and the phase of state $|1\rangle_{1}|1\rangle_{2}$ (magenta) for
the system initially in $|d_{2}(0)\rangle$. Other parameters as same in Fig.
\ref{fig:population_geo}. }
\end{figure}

\textbf{Scheme 2.} Gate based on non-Berry adiabatic phase arisen
from staircase phase control. The operation procedure is generally
divided into two steps during the time interval $0\leqslant t\leqslant2\tau$,
in which the time-dependent amplitudes of the Rabi frequencies again
vary according to Eq.(\ref{eq:pulse_s1}), and the phases of the driving
fields follow
\begin{equation}
\phi_{p}(t)=\textrm{const.},\:\phi_{r}(t)=\frac{\pi}{2}\Theta(t-\tau),
\end{equation}
with $\Theta(x)$ being the unit step function. Note that the relative
phase is changed only at the end of the first half of the pulse sequence
($t=\tau$) without the limit of adiabaticity, and the system Hamiltonian
is not changed along a closed curve in the parameter space $\mathbf{R}(t)=(\theta(t),\phi_{r}(t))$.
Therefore, it is fundamentally different from the geometric operation
(leading to the Berry phase) proposed by Møller et al. \textcolor{black}{\cite{Moller2008}},
where the relative phase should be adiabatically modulated when the
applied pulses overlap, and the initial and the final Hamiltonian
of the evolutional system should remain the same (i.e. $\mathbf{R}(2\tau)=\mathbf{R}(0)$). The idea of realizing
a phase gate through adiabatic manipulation of the dark state with
staircase phase control was firstly studied in ion traps \cite{Zheng_PRL2005}.

In the first step ($0\leqslant t\leqslant\tau$), the phase factors
$\phi_{p}$, $\phi_{\mu}$ are set to be equal so that $\phi_{r}=0$,
e.g. $\phi_{p}=\phi_{\mu}=0$ for simplicity. $\theta$ is adiabatically
increased from $0$ to $\pi/2$ by adjusting the relative intensity
of the coupling fields as in Eq. (\ref{eq:pulse_s1}). For $V_{22}\ll V_{23,}V_{33}$,
the temporal evolution of the basis states will follow the dark states
{[}Eqs. (\ref{eq:ds2_norm}) and (\ref{eq:ds1}) {]} throughout the
procedure, leading to the transformations
\begin{gather}
|0\rangle_{1}|0\rangle_{2}\rightarrow|0\rangle_{1}|0\rangle_{2},\quad|0\rangle_{1}|1\rangle_{2}\rightarrow-|0\rangle_{1}|3\rangle_{2},\nonumber \\
|1\rangle_{1}|0\rangle_{2}\rightarrow-|3\rangle_{1}|0\rangle_{2},\quad|1\rangle_{1}|1\rangle_{2}\rightarrow\frac{1}{\sqrt{2}}(-|\phi_{1}\rangle+|\phi_{4}\rangle).
\end{gather}

In the second step ($\tau\leqslant t\leqslant2\tau$), $\theta$ is
tuned adiabatically from $\pi/2$ back to $0$ but with $\phi_{p}=0$
and $\phi_{\mu}=\pi/2$ (i.e. $\phi_{r}=\pi/2$), which gives
rise to
\begin{gather}
|0\rangle_{1}|0\rangle_{2}\rightarrow|0\rangle_{1}|0\rangle_{2},\quad-|0\rangle_{1}|3\rangle_{2}\rightarrow e^{i\pi/2}|0\rangle_{1}|1\rangle_{2},\nonumber \\
-|3\rangle_{1}|0\rangle_{2}\rightarrow e^{i\pi/2}|1\rangle_{1}|0\rangle_{2},\quad\frac{1}{\sqrt{2}}(-|\phi_{1}\rangle+|\phi_{4}\rangle)\rightarrow|1\rangle_{1}|1\rangle_{2}.\label{eq:Ng_transf}
\end{gather}
Since the two processes are highly adiabatic, the population of the
basis states return to the initial state after the counterintuitive
pulse sequence. It is interesting to see that the basis states $|0\rangle_{1}|1\rangle_{2}$
and $|1\rangle_{1}|0\rangle_{2}$ finally acquire an additional phase
factor $e^{i\varphi_{ng}}$ with $\varphi_{ng}=\text{\ensuremath{\pi}/2}$,
which does not exist for $|0\rangle_{1}|0\rangle_{2}$ and $|1\rangle_{1}|1\rangle_{2}$
(see Fig. \ref{fig:population_ng}). Because the dark states are
the eigenstates of $\mathcal{H}_{i}$ ($i=1,2$) with zero eigenvalues, $\varphi_{ng}$
has no dynamic origin. On the other hand, here the Hamiltonian is
not required to make a cyclic evolution in the parameter space as for
the accumulation of the adiabatic Berry phase. 

Finally, by applying single-qubit operations $|1\rangle_{1,2}\rightarrow e^{i\pi/2}|1\rangle_{1,2}$
to both atoms, we recover the familiar controlled-Z gate
\begin{gather}
|0\rangle_{1}|0\rangle_{2}\rightarrow|0\rangle_{1}|0\rangle_{2},\quad|0\rangle_{1}|1\rangle_{2}\rightarrow|0\rangle_{1}|1\rangle_{2},\nonumber \\
|1\rangle_{1}|0\rangle_{2}\rightarrow|1\rangle_{1}|0\rangle_{2},\quad|1\rangle_{1}|1\rangle_{2}\rightarrow-|1\rangle_{1}|1\rangle_{2},\label{eq:QPG}
\end{gather}
which can be easily transformed to a controlled-NOT gate by using
two additional $\pi/2$ pulses rotating the target qubit around the
$y$-axis in the opposite directions. Note that for $V_{33}\ll V_{22,}V_{23}$,
repeating the operation procedure above will lead to the transformation
for the basis states: $|0\rangle_{1}|1\rangle_{2}\rightarrow e^{i\pi/2}|0\rangle_{1}|1\rangle_{2}$,
$|1\rangle_{1}|0\rangle_{2}\rightarrow e^{i\pi/2}|1\rangle_{1}|0\rangle_{2}$
and $e^{i\pi}|1\rangle_{1}|1\rangle_{2}$, which is impossible to
become a universal binary gate under any local operations. 

From a comparison between the two schemes we can see that the non-Berry
phase gate via the staircase phase control is built on a completely
different mechanism in contrast to the normal dynamical and geometric
phase gates: the qubit system does not undergo any dynamical phase
shift since it works in the zero-energy eigenspace; the Hamiltonian
is not changed along a closed curve in the parameter space; precisely
adiabatic modulation of the phases of the driving fields and adiabatic
control of the population transfer at the same time is unnecessary,
thus, the errors in obtaining the required geometric solid angle are
avoided and the operation procedure is simplified.

\section{Physical realization: asymmetric Rydberg coupling}

In the context of Rydberg experiments, the strongly asymmetric coupling
condition $V_{22}\ll V_{23,}V_{33}$ can be found, for example, by
mapping the Rydberg states to $|2=40p_{3/2},m=1/2\rangle$ and $|3=41s_{1/2},m=1/2\rangle$
of Rubidium atoms separated at an interatomic distance $R$ of several
micrometers. In this case, the blockade interaction between the states
$|2\rangle$ and $|3\rangle$ is an exchange process of resonant dipole
nature ($\sim n^{4}/R^{3}$ with $n$ being the principal quantum
number), where the zero-interaction angle can be avoided either by
using a spatial light modulator to create the preset trap pattern
or by applying a weak external magnetic field ($B=10^{-7}$ T) to
couple the atomic Zeeman states of different magnetic quantum numbers.
The anisotropic interaction between states $|2\rangle$ and isotropic
interaction between states $|3\rangle$ are both induced by the Förster
process, where the two-atomic interaction potential can transit from
the dipole-dipole to the van der Waals limit ($\sim n^{11}/R^{6}$),
depending on the interatomic distance \cite{Saffman2008}. It is therefore
possible to restrict our consideration to the asymmetric coupling
regime, which represents the dominant interaction mechanism at the
atomic separation of interest. For $R=3$$\mu$m, the interaction
strengths $V_{23}$ and $V_{22}$ can respectively vary from $5$
MHz to $20$ MHz, and from $0.02$ MHz to $0.1$ MHz by adjusting
the angle between the dipoles, and the interaction strength $V_{33}$
approximates $2\pi\times3.7$ MHz \cite{PhysRevLett.102.240502}. 

On the other hand, the excitation of Rydberg $p$-states from ground
$s$-states in a single photon transition has recently become feasible
due to the availability of ultraviolet (UV) laser sources, which results
in much larger Rabi frequency $\Omega$ (scaling as $\Omega\sim n^{-3/2}$)
compared to a three-photon excitation process \cite{Hankin2014b}.
In addition, the optical excitation of a Rydberg state followed by
a microwave-driven coupling between two neighboring Rydberg levels
has been experimentally demonstrated as well, where the Rabi frequency
of the Rydberg-Rydberg transition can reach several tens of MHz by
increasing the intensity of the microwave field \cite{PhysRevLett.110.103001}.
Thus, it becomes very promising to implement the proposed schemes
with asymmetric Rydberg-Rydberg interaction by integrating the current
experimental techniques \cite{Barredo2015}.

\begin{figure}
\begin{centering}
\includegraphics[width=0.9\columnwidth]{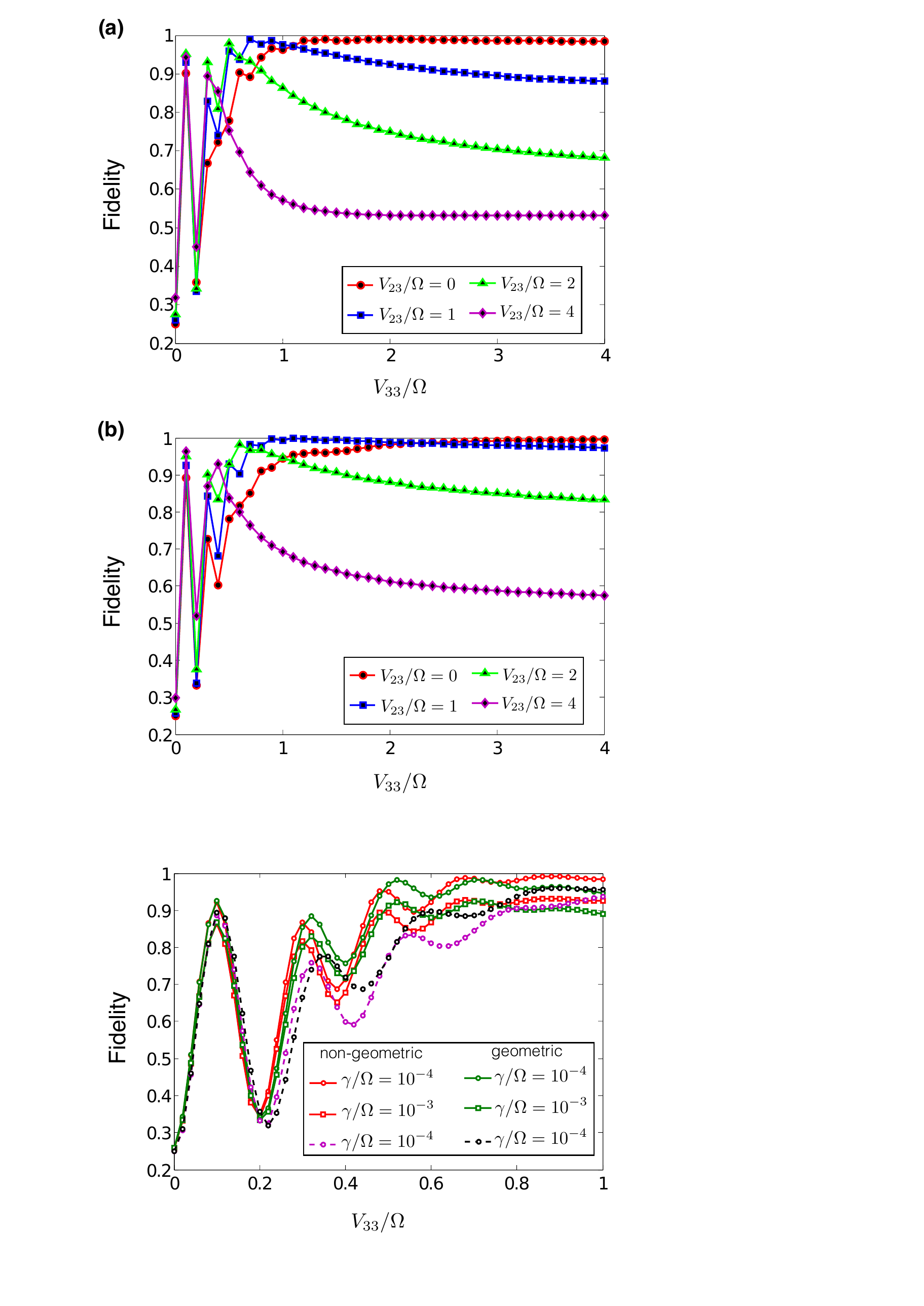}
\par\end{centering}
\caption{\label{fig:fidelity_Nsp}(color online) Fidelities of gates based
on Berry phase (a) and non-Berry adiabatic phase (b) v.s. the energy
shift $V_{33}/\Omega$ of the collective Rydberg states $|3\rangle_{1}|3\rangle_{2}$
for the DDI strength $V_{23}/\Omega=0,1,2\text{,4}$ (from top to
bottom) and $V_{22}/\Omega=0.005$. Other parameters are as in Fig.
\ref{fig:population_geo}.}
\end{figure}

To evaluate the performance of the controlled-Z gate, we use the fidelity
$F=[\textrm{Tr}\sqrt{\sqrt{\rho_{tar}}\rho(2\tau)\sqrt{\rho_{tar}}}]^{2}$
to measure the desired output $\rho_{tar}$ given an input of all
the logical states $|\psi_{0}\rangle=\frac{1}{2}(|0\rangle_{1}|0\rangle_{2}+|0\rangle_{1}|1\rangle_{2}+|1\rangle_{1}|0\rangle_{2}+|1\rangle_{1}|1\rangle_{2})$,
where $\rho_{tar}=|\psi_{tar}\rangle\langle\psi_{tar}|$ with $|\psi_{tar}\rangle$
being the target state obtained through an ideal gate operation $|\psi_{tar}\rangle=U_{CZ}|\psi_{0}\rangle$,
and $\rho(2\tau)$ is the actual output state in the logical space
produced in the presence of the error sources, such as nonadiabatic
transitions, docoherence induced by atomic spontaneous emission and
atomic motion. In Fig. \ref{fig:fidelity_Nsp}, we have shown the
fidelity of the controlled-Z gate {[} Eq.(\ref{eq:geo-gate}) and
Eq.(\ref{eq:QPG}){]} under the condition of asymmetric Rydberg coupling
$V_{22}\ll V_{23},V_{33}$ in the coherent regime. Considering \textcolor{black}{the
dynamically perturbative effect of $V_{22}$} (\textcolor{black}{i.e.
$V_{22}\tau\ll1$}), the fidelity reaches its optimum at $V_{33}=0.7$, $V_{23}=1$ and  $V_{33}=0.9$, $V_{23}=1$ for gates based on the Berry 
and the non-Berry adiabatic phase, respectively, where the interaction
strengths $V_{23},V_{33}$ are of comparable magnitude with the maximum
of the Rabi frequencies $\Omega$, corresponding to the intermediate
coupling regime $V_{23},V_{33}\sim\Omega$. Further increasing $V_{23}$
or $V_{33}$ will lead to reduction of the gate fidelity (due to nonadiabatic
transfer towards the nonzero-energy eigenstates), however, note that
the non-Berry operation is more robust against the variation of the
Rydberg interactions compared with the geometric Berry operation.
For the special situation where $V_{22}\simeq0$ and $V_{23}=0$,
available for a cascaded level configuration involving a single Rydberg
state (see later discussion), the condition for a high-fidelity gate
performance is simply $V_{33}>2\Omega$, which lies in the regime
of Rydberg blockade. In this case, the optimal implementation of the
Berry-phase-based controlled-Z gate requires slightly weaker $V_{33}$
than that for the non-Berry adiabatic operation, but again, the latter
exhibits its robustness as $V_{33}$ increases. 

\begin{figure}
\begin{centering}
\includegraphics[width=0.9\columnwidth]{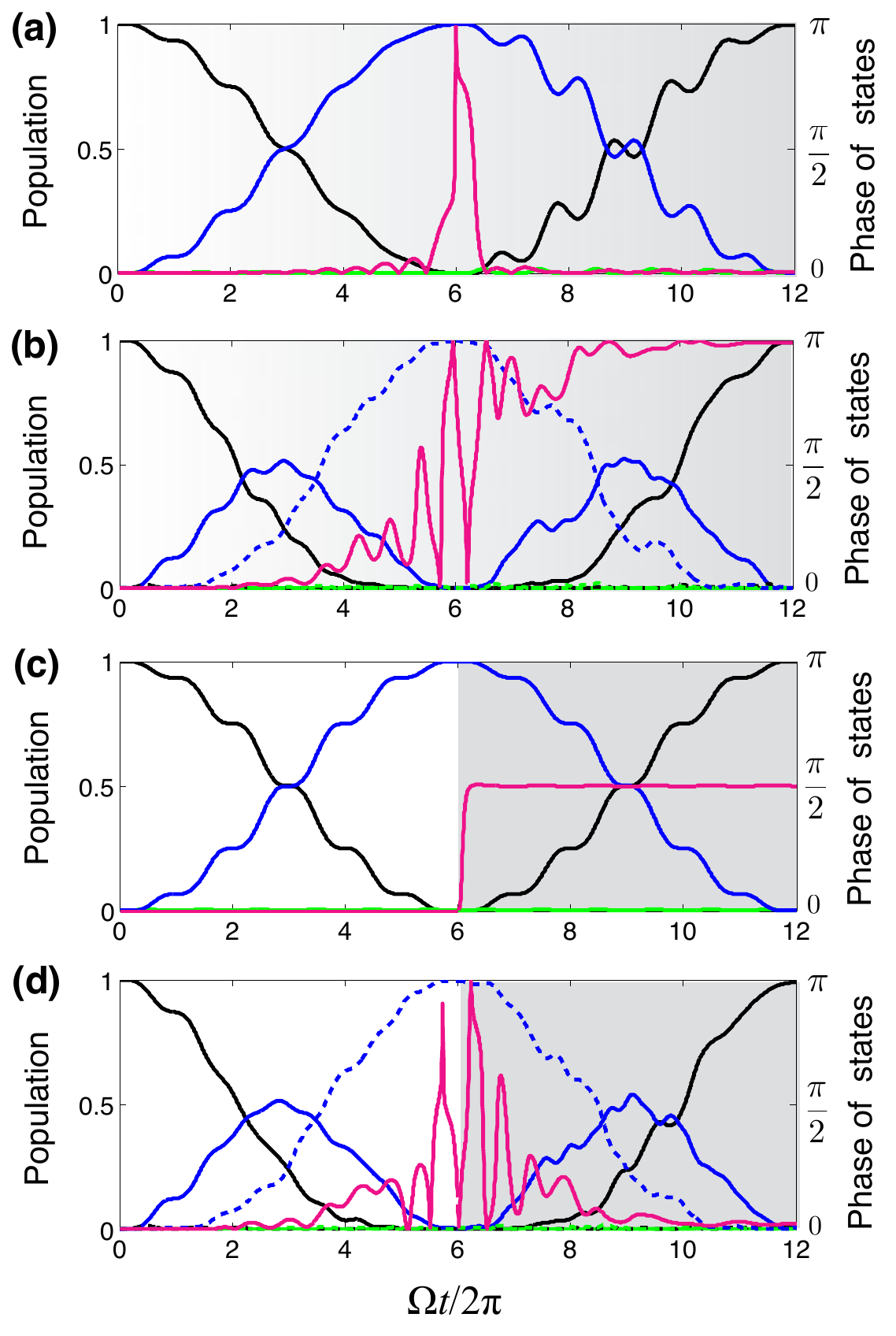}
\par\end{centering}
\caption{\label{fig:population_DynPG}(color online) Time dependence of the
state populations during the gate operations based on Berry phase
{[}(a), (b){]} and non-Berry adiabatic phase {[}(c), (d){]} with $(V_{22},V_{23},V_{33})/\Omega=(1,1.5,0.1)$.
The color scheme and other parameters are as in Fig. \ref{fig:population_geo}.}
\end{figure}

For small $V_{33}$, the temporal evolution of the system is no longer
adiabatically confined in the state $|d_{2}(t)\rangle$ and the effect
of the other dark component $|d_{2}'(t)\rangle$ should be considered.
In this case, the nonadiabatic transition to the doubly excited state
$|3\rangle_{1}|3\rangle_{2}$ accompanied with interatomic interaction
will introduce a dynamical phase, which may be constructive for implementing
the controlled-Z gate as well. To gain the insight, we have repeated
the procedures for generating the Berry and the non-Berry phases as before
under the condition of $V_{33}\ll V_{22},V_{23}$. If $V_{33}=0$,
the system strictly evolves along the dark state $|d'_{2}(t)\rangle$,
where the phase difference $\phi_{r}$ of the control fields becomes
the only relevant phase factor for the modulation process. For the
operation to obtain Berry phases, the system acquires no geometric
phase during the cyclic evolution since $\phi_{r}$ is kept invariant
{[}see Fig. \ref{fig:population_DynPG}(a-b){]}. Alternatively, for
the operation to obtain non-Berry adiabatic phases, the rise up of
$\phi_{r}$ at $t=\tau$ introduces phase factors $e^{i2\phi_{r}}=e^{i\pi}$
and $e^{i\phi_{r}}=e^{i\pi/2}$ to the basis states $|1\rangle_{1}|1\rangle_{2}$
and $|1\rangle_{1}|0\rangle_{2}$ (or $|0\rangle_{1}|1\rangle_{2}$),
respectively, which are irrelevant to a binary gate. While for a finite
$V_{33}$, the instantaneous ground state of $\mathcal{H}'_{R}$ evolves
from the bare $|1\rangle_{1}|1\rangle_{2}$ state into a ``dressed''
state with some admixture of $|3\rangle_{1}|3\rangle_{2}$, which
additionally supplements a dynamical phase $\varphi_{d}\approx\int\textrm{sin}^{4}\theta V_{33}dt$
to $|1\rangle_{1}|1\rangle_{2}$ {[}see Fig. \ref{fig:population_DynPG}(c-d){]}.
Therefore, the implementation of a controlled-Z gate via the completely
dynamical control is still available for $\varphi_{d}=\pi$ for both
cases, and is sensitive to the fluctuation of Rydberg interactions
nevertheless.

\section{The effect of Spontaneous emission and interatomic force}

\begin{figure}
\begin{centering}
\includegraphics[width=0.9\columnwidth]{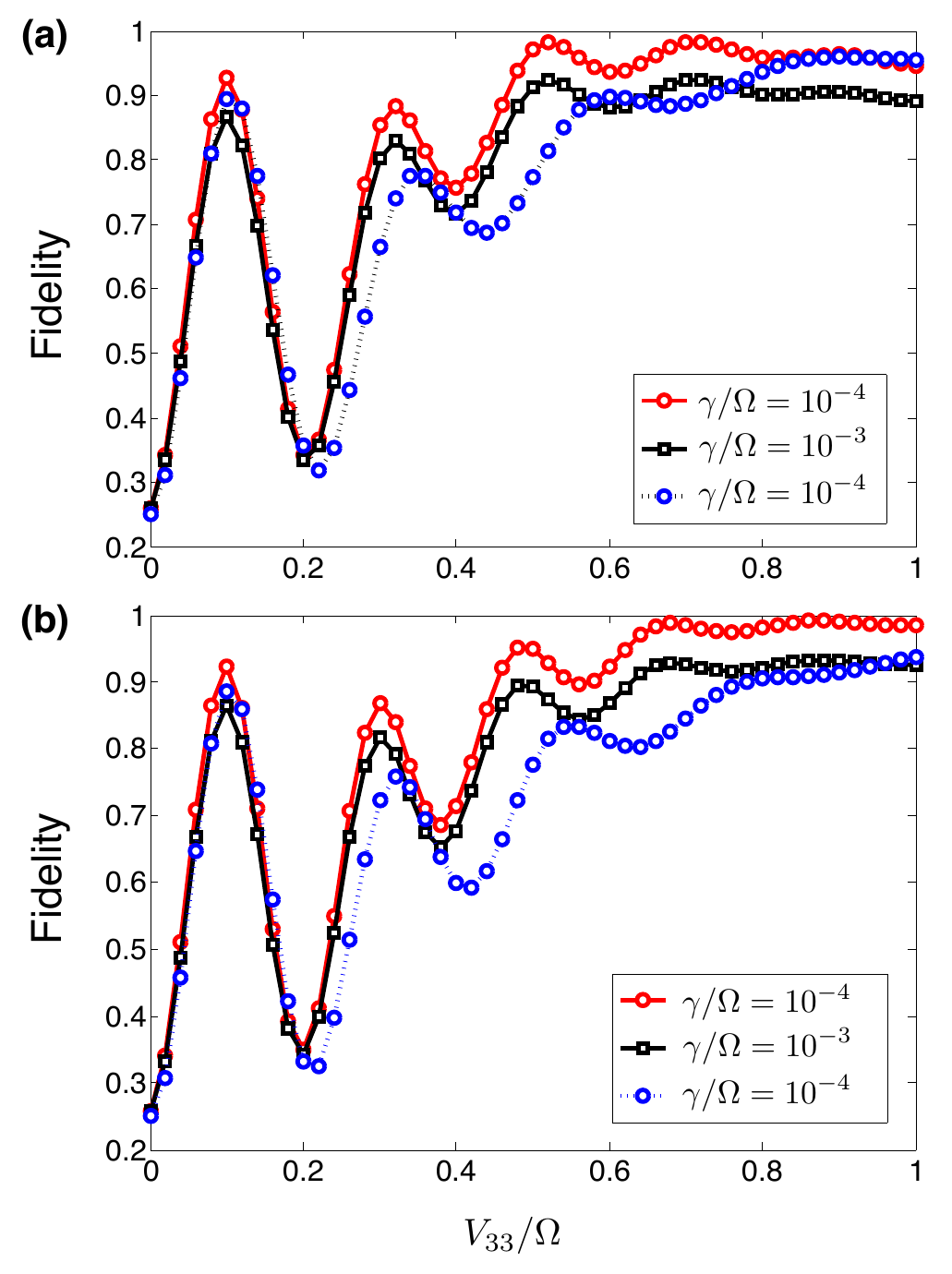}
\par\end{centering}
\caption{\label{fig:Fide_sp}(color online) Fidelities of the gates based on
Berry phase (a) and non-Berry adiabatic phase (b) for different atomic
spontaneous emission rates as functions of the Rydberg interaction
$V_{33}$ for \textcolor{black}{$(V_{22,}V_{23})/\Omega=(0.005,1)$}
(solid) and \textcolor{black}{$V_{22}=V_{23}=0$} (dash). We here
set $\gamma_{2}=\gamma_{3}=\gamma$.}
\end{figure}

The two atoms excited to Rydberg states are subjected to decoherence
due to atomic spontaneous emission and interatomic force. The dissipative
dynamics can be calculated by the Lindblad master equation for the
density operator $\rho$ of the two-atom system,
\begin{eqnarray}
\dot{\rho}(t) & = & -i[\mathcal{H}'_{R},\rho(t)]+\sum_{i=1}^{2}\sum_{k=1}^{2}\mathcal{L}[A_{i,k}]\rho(t),\label{eq:master_eq}
\end{eqnarray}
where $\mathcal{L}[A_{i,k}]\rho=A_{i,k}\rho A_{i,k}^{\dagger}-\frac{1}{2}\{A_{i,k}^{\dagger}A_{i,k},\rho\}$,
$A_{i,1}=\sqrt{\gamma_{2}}|1\rangle_{ii}\langle2|$ and $A_{i,2}=\sqrt{\gamma_{3}}|2\rangle_{ii}\langle3|$
with $\gamma_{2}$ and $\gamma_{3}$ being the spontaneous decay rates
for the transition channels $|2\rangle_{i}\rightarrow|1\rangle_{i}$
and $|3\rangle_{i}\rightarrow|2\rangle_{i}$ respectively. In Fig.
\ref{fig:Fide_sp}, we show overlap (fidelity) between the realistic
density matrix $\rho(2\tau)$ at the end of the pulse sequences from
Eq.(\ref{eq:master_eq}) and the ideal result $\rho_{tar}$ (for $\gamma_{2}=\gamma_{3}=0$),
for the system initially in $|\psi_{0}\rangle$. Since the lifetime
of the Rydberg states $|2\rangle$ ($|3\rangle$) with principal quantum
number $n=40$ or $41$ is around $60\mu s$, thus the decay rates
are taken as $\gamma_{2}=\gamma_{3}=10^{-4}\Omega$, $10^{-3}\Omega$,
corresponding to the peak Rabi frequency $\Omega/2\pi=20$MHz, $2$MHz,
respectively. For the former case, the fidelities of the gates relying
on the Berry phase and non-Berry adiabatic phase are 0.983 and 0.992,
respectively; while for the latter case, fidelity of better than 0.92
is still achievable for both schemes. On the other hand, the interatomic
force (induced by double excitation of Rydberg states) during the
gate operation can couple the internal degree of freedom to the externally
atomic motion. Its perturbative effect on the gate fidelity can be
estimated by $\sim\frac{3\lambda_{0}V_{33}}{R\omega_{0}}(1-e^{-i\omega_{0}\tau})$
to the first order, with $\omega_{0}$ the trapping frequency and
$\lambda_{0}$ the wavelength of trapping light \cite{Rao2014}. Thus,
one can enlarge the Rabi frequency $\Omega$ to reduce the gate duration
$\tau$ or alternatively use an optical lattice (instead of an optical
tweezer trap) with higher trapping frequency to trap the atoms such
that the motional effect can be reasonably ignored.

\begin{figure}
\begin{centering}
\includegraphics[width=0.8\columnwidth]{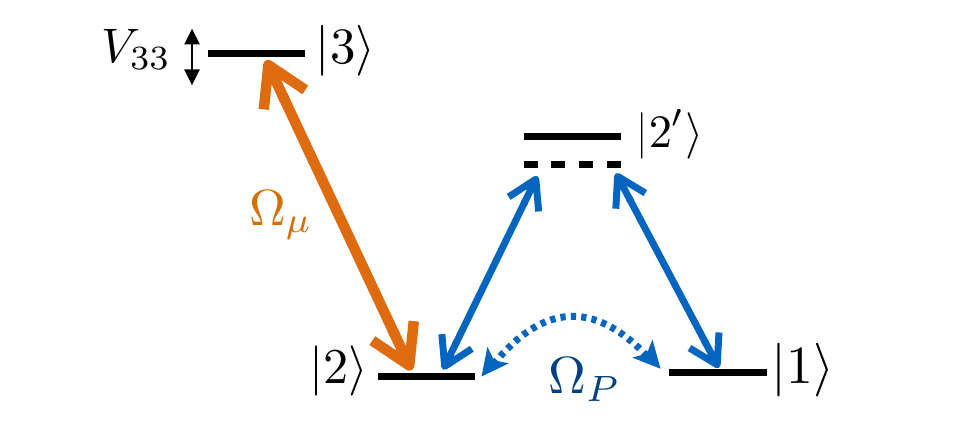}
\par\end{centering}
\caption{\label{fig:level_Raman}(color online) Level configuration. Two hyperfine
ground states $|1\rangle$, $|2\rangle$ are coupled via the Raman
process with the effective Rabi frequency $\Omega_{p}$. The Rydberg
state $|3\rangle$ is excited via a single photon transition of the
Rabi frequency $\Omega_{\mu}$.}
\end{figure}

As mentioned before, the controlled-Z gate can be implemented as well
with the atomic level scheme involving only a single Rydberg state.
In this case, the microwave control becomes unnecessary. However, for
the usual Rydberg EIT configuration (i.e. a cascaded three-level system),
the populated intermediate state $|2\rangle$ (such as $6p_{1/2}$
for Rb and $7p_{1/2}$ for Cs atoms) is an excited state with strong
spontaneous emission rate, which will irreversibly deteriorate the
coherent population transfer and then the gate fidelity \cite{Moller2008,Rao2014}.
However, the obstacle can be overcome by using the single-photon excitation
scheme for the ground-Rydberg transition and by mapping $|1\rangle$
and $|2\rangle$ to the atomic hyperfine states, which can couple
to each other via two-photon Raman processes (see Fig. \ref{fig:level_Raman}).
Therefore, the fidelity can be further improved by selecting a Rydberg
state with larger principal quantum number and longer lifetime. For
example, in a 300K environment, the Cs Rydberg states $|90p\rangle$,
$|95p\rangle$ have the lifetimes $361\text{\ensuremath{\mu}s}$ and
$406\text{\ensuremath{\mu}s}$, respectively \cite{Beterov2016a}.
Moreover, double excitation of Rydberg states is avoided, and thus
the effect of the interatomic force that may entangle their motional
degree of freedom can be neglected.

\section{Conclusion}

In conclusion, we have shown that the Rydberg-Rydberg interaction
between two highly excited atoms can be exploited for implementing
a reliable controlled-Z gate via adiabatic passage and tailored phase
modulation. The newly developed addressing schemes drive the system
Hamiltonian to change in a cyclic or a noncyclic manner, giving rise
to a Berry phase or a non-Berry adiabatic phase for implementation
of conditional phase gates. In the former case, the geometric phase
is acquired through concurrent control of the phases of the driving
fields, and can be alternatively obtained via modulation of the relative
phase in two different ways, while for the latter the requirement
of adiabatic phase control becomes unnecessary and therefore the experimental
complexity can be significantly released. We also pointed out that
the implementation of the schemes with multilevel atomic configuration
involving a unique Rydberg state might be more promising for experimental
demonstration. We note that our adiabatic Rydberg gates may not replace
the conventional approaches with fast dynamical control, however,
the merits of the adiabatic technique itself and the new addressing
schemes of phase modulation found here will provide new perspectives
for adiabatic manipulation of interacting Rydberg systems.

\section*{Acknowledgments}

This work was supported by the National Natural Science Foundation
of China under Grants No. 11374054, No. 11534002, No. 11674060, No.
11575045, and No. 11405031; and the Natural Science Foundation of
Fujian Province under Grant No. 2017J01401.

\bibliographystyle{apsrev}
\bibliography{Ryd_gates,Ryd_QIP,Ryd_exp,Ryd_AP,geo_zheng,others,Ryd_asym}

\begin{thebibliography}{46}
\expandafter\ifx\csname natexlab\endcsname\relax\def\natexlab#1{#1}\fi
\expandafter\ifx\csname bibnamefont\endcsname\relax
  \def\bibnamefont#1{#1}\fi
\expandafter\ifx\csname bibfnamefont\endcsname\relax
  \def\bibfnamefont#1{#1}\fi
\expandafter\ifx\csname citenamefont\endcsname\relax
  \def\citenamefont#1{#1}\fi
\expandafter\ifx\csname url\endcsname\relax
  \def\url#1{\texttt{#1}}\fi
\expandafter\ifx\csname urlprefix\endcsname\relax\def\urlprefix{URL }\fi
\providecommand{\bibinfo}[2]{#2}
\providecommand{\eprint}[2][]{\url{#2}}

\bibitem[{\citenamefont{Saffman et~al.}(2010)\citenamefont{Saffman, Walker, and
  M{\o}lmer}}]{Saffman2010}
\bibinfo{author}{\bibfnamefont{M.}~\bibnamefont{Saffman}},
  \bibinfo{author}{\bibfnamefont{T.~G.} \bibnamefont{Walker}},
  \bibnamefont{and}
  \bibinfo{author}{\bibfnamefont{K.}~\bibnamefont{M{\o}lmer}},
  \bibinfo{journal}{Rev. Mod. Phys.} \textbf{\bibinfo{volume}{82}},
  \bibinfo{pages}{2313} (\bibinfo{year}{2010}).

\bibitem[{\citenamefont{Browaeys et~al.}(2016)\citenamefont{Browaeys, Barredo,
  and Lahaye}}]{Browaeys2016}
\bibinfo{author}{\bibfnamefont{A.}~\bibnamefont{Browaeys}},
  \bibinfo{author}{\bibfnamefont{D.}~\bibnamefont{Barredo}}, \bibnamefont{and}
  \bibinfo{author}{\bibfnamefont{T.}~\bibnamefont{Lahaye}},
  \bibinfo{journal}{J. Phys. B: At. Mol. Opt. Phys.}
  \textbf{\bibinfo{volume}{49}}, \bibinfo{pages}{152001}
  (\bibinfo{year}{2016}).

\bibitem[{\citenamefont{Cozzini et~al.}(2006)\citenamefont{Cozzini, Calarco,
  Recati, and Zoller}}]{Cozzini2006375}
\bibinfo{author}{\bibfnamefont{M.}~\bibnamefont{Cozzini}},
  \bibinfo{author}{\bibfnamefont{T.}~\bibnamefont{Calarco}},
  \bibinfo{author}{\bibfnamefont{A.}~\bibnamefont{Recati}}, \bibnamefont{and}
  \bibinfo{author}{\bibfnamefont{P.}~\bibnamefont{Zoller}},
  \bibinfo{journal}{Opt. Commun.} \textbf{\bibinfo{volume}{264}},
  \bibinfo{pages}{375} (\bibinfo{year}{2006}).

\bibitem[{\citenamefont{Brion et~al.}(2007)\citenamefont{Brion, Pedersen, and
  M{\o}lmer}}]{Brion2007}
\bibinfo{author}{\bibfnamefont{E.}~\bibnamefont{Brion}},
  \bibinfo{author}{\bibfnamefont{L.~H.} \bibnamefont{Pedersen}},
  \bibnamefont{and}
  \bibinfo{author}{\bibfnamefont{K.}~\bibnamefont{M{\o}lmer}},
  \bibinfo{journal}{J. Phys. B: At. Mol. Opt. Phys.}
  \textbf{\bibinfo{volume}{40}}, \bibinfo{pages}{S159} (\bibinfo{year}{2007}).

\bibitem[{\citenamefont{M{\"{u}}ller et~al.}(2009)\citenamefont{M{\"{u}}ller,
  Lesanovsky, Weimer, B{\"{u}}chler, and Zoller}}]{Muller2009}
\bibinfo{author}{\bibfnamefont{M.}~\bibnamefont{M{\"{u}}ller}},
  \bibinfo{author}{\bibfnamefont{I.}~\bibnamefont{Lesanovsky}},
  \bibinfo{author}{\bibfnamefont{H.}~\bibnamefont{Weimer}},
  \bibinfo{author}{\bibfnamefont{H.~P.} \bibnamefont{B{\"{u}}chler}},
  \bibnamefont{and} \bibinfo{author}{\bibfnamefont{P.}~\bibnamefont{Zoller}},
  \bibinfo{journal}{Phys. Rev. Lett.} \textbf{\bibinfo{volume}{102}},
  \bibinfo{pages}{170502} (\bibinfo{year}{2009}).

\bibitem[{\citenamefont{Wu et~al.}(2010)\citenamefont{Wu, Yang, and
  Zheng}}]{PhysRevA.82.034307}
\bibinfo{author}{\bibfnamefont{H.-Z.} \bibnamefont{Wu}},
  \bibinfo{author}{\bibfnamefont{Z.-B.} \bibnamefont{Yang}}, \bibnamefont{and}
  \bibinfo{author}{\bibfnamefont{S.-B.} \bibnamefont{Zheng}},
  \bibinfo{journal}{Phys. Rev. A} \textbf{\bibinfo{volume}{82}},
  \bibinfo{pages}{034307} (\bibinfo{year}{2010}).

\bibitem[{\citenamefont{Rao and M{\o}lmer}(2014)}]{Rao2014}
\bibinfo{author}{\bibfnamefont{D.~D.~B.} \bibnamefont{Rao}} \bibnamefont{and}
  \bibinfo{author}{\bibfnamefont{K.}~\bibnamefont{M{\o}lmer}},
  \bibinfo{journal}{Phys. Rev. A} \textbf{\bibinfo{volume}{89}},
  \bibinfo{pages}{030301} (\bibinfo{year}{2014}).

\bibitem[{\citenamefont{Keating et~al.}(2015)\citenamefont{Keating, Cook,
  Hankin, Jau, Biedermann, and Deutsch}}]{Keating2015}
\bibinfo{author}{\bibfnamefont{T.}~\bibnamefont{Keating}},
  \bibinfo{author}{\bibfnamefont{R.~L.} \bibnamefont{Cook}},
  \bibinfo{author}{\bibfnamefont{A.~M.} \bibnamefont{Hankin}},
  \bibinfo{author}{\bibfnamefont{Y.-Y.} \bibnamefont{Jau}},
  \bibinfo{author}{\bibfnamefont{G.~W.} \bibnamefont{Biedermann}},
  \bibnamefont{and} \bibinfo{author}{\bibfnamefont{I.~H.}
  \bibnamefont{Deutsch}}, \bibinfo{journal}{Phys. Rev. A}
  \textbf{\bibinfo{volume}{91}}, \bibinfo{pages}{012337}
  (\bibinfo{year}{2015}).

\bibitem[{\citenamefont{Beterov et~al.}(2016)\citenamefont{Beterov, Saffman,
  Yakshina, Tretyakov, Entin, Bergamini, Kuznetsova, and
  Ryabtsev}}]{Beterov2016a}
\bibinfo{author}{\bibfnamefont{I.~I.} \bibnamefont{Beterov}},
  \bibinfo{author}{\bibfnamefont{M.}~\bibnamefont{Saffman}},
  \bibinfo{author}{\bibfnamefont{E.~A.} \bibnamefont{Yakshina}},
  \bibinfo{author}{\bibfnamefont{D.~B.} \bibnamefont{Tretyakov}},
  \bibinfo{author}{\bibfnamefont{V.~M.} \bibnamefont{Entin}},
  \bibinfo{author}{\bibfnamefont{S.}~\bibnamefont{Bergamini}},
  \bibinfo{author}{\bibfnamefont{E.~A.} \bibnamefont{Kuznetsova}},
  \bibnamefont{and} \bibinfo{author}{\bibfnamefont{I.~I.}
  \bibnamefont{Ryabtsev}}, \bibinfo{journal}{Phys. Rev. A}
  \textbf{\bibinfo{volume}{94}}, \bibinfo{pages}{062307}
  (\bibinfo{year}{2016}).

\bibitem[{\citenamefont{Zeiher et~al.}(2015)\citenamefont{Zeiher, Schau{\ss},
  Hild, Macr{\`{i}}, Bloch, and Gross}}]{Zeiher2015}
\bibinfo{author}{\bibfnamefont{J.}~\bibnamefont{Zeiher}},
  \bibinfo{author}{\bibfnamefont{P.}~\bibnamefont{Schau{\ss}}},
  \bibinfo{author}{\bibfnamefont{S.}~\bibnamefont{Hild}},
  \bibinfo{author}{\bibfnamefont{T.}~\bibnamefont{Macr{\`{i}}}},
  \bibinfo{author}{\bibfnamefont{I.}~\bibnamefont{Bloch}}, \bibnamefont{and}
  \bibinfo{author}{\bibfnamefont{C.}~\bibnamefont{Gross}},
  \bibinfo{journal}{Phys. Rev. X} \textbf{\bibinfo{volume}{5}},
  \bibinfo{pages}{031015} (\bibinfo{year}{2015}).

\bibitem[{\citenamefont{Brion et~al.}(2008)\citenamefont{Brion, Pedersen,
  Saffman, and M$\backslash$olmer}}]{PhysRevLett.100.110506}
\bibinfo{author}{\bibfnamefont{E.}~\bibnamefont{Brion}},
  \bibinfo{author}{\bibfnamefont{L.~H.} \bibnamefont{Pedersen}},
  \bibinfo{author}{\bibfnamefont{M.}~\bibnamefont{Saffman}}, \bibnamefont{and}
  \bibinfo{author}{\bibfnamefont{K.}~\bibnamefont{M{\o}lmer}},
  \bibinfo{journal}{Phys. Rev. Lett.} \textbf{\bibinfo{volume}{100}},
  \bibinfo{pages}{110506} (\bibinfo{year}{2008}).

\bibitem[{\citenamefont{Chen}(2011)}]{Chen2011}
\bibinfo{author}{\bibfnamefont{A.}~\bibnamefont{Chen}}, \bibinfo{journal}{Opt.
  Express} \textbf{\bibinfo{volume}{19}}, \bibinfo{pages}{2037}
  (\bibinfo{year}{2011}).

\bibitem[{\citenamefont{Sanders et~al.}(2014)\citenamefont{Sanders, van Bijnen,
  Vredenbregt, and Kokkelmans}}]{Sanders2014}
\bibinfo{author}{\bibfnamefont{J.}~\bibnamefont{Sanders}},
  \bibinfo{author}{\bibfnamefont{R.}~\bibnamefont{van Bijnen}},
  \bibinfo{author}{\bibfnamefont{E.}~\bibnamefont{Vredenbregt}},
  \bibnamefont{and}
  \bibinfo{author}{\bibfnamefont{S.}~\bibnamefont{Kokkelmans}},
  \bibinfo{journal}{Phys. Rev. Lett.} \textbf{\bibinfo{volume}{112}},
  \bibinfo{pages}{163001} (\bibinfo{year}{2014}).

\bibitem[{\citenamefont{Petrosyan et~al.}()\citenamefont{Petrosyan, Saffman,
  and M{\o}lmer}}]{Petrosyan}
\bibinfo{author}{\bibfnamefont{D.}~\bibnamefont{Petrosyan}},
  \bibinfo{author}{\bibfnamefont{M.}~\bibnamefont{Saffman}}, \bibnamefont{and}
  \bibinfo{author}{\bibfnamefont{K.}~\bibnamefont{M{\o}lmer}},
  \bibinfo{journal}{J. Phys. B: At. Mol. Opt. Phys.}
  \textbf{\bibinfo{volume}{49}}, \bibinfo{pages}{094004} (2016).

\bibitem[{\citenamefont{Han et~al.}(2010)\citenamefont{Han, He, Heshami, Li,
  and Simon}}]{Han2010}
\bibinfo{author}{\bibfnamefont{Y.}~\bibnamefont{Han}},
  \bibinfo{author}{\bibfnamefont{B.}~\bibnamefont{He}},
  \bibinfo{author}{\bibfnamefont{K.}~\bibnamefont{Heshami}},
  \bibinfo{author}{\bibfnamefont{C.~Z.} \bibnamefont{Li}}, \bibnamefont{and}
  \bibinfo{author}{\bibfnamefont{C.}~\bibnamefont{Simon}},
  \bibinfo{journal}{Phys. Rev. A}
  \textbf{\bibinfo{volume}{81}}, \bibinfo{pages}{052311} (\bibinfo{year}{2010}).

\bibitem[{\citenamefont{Zhao et~al.}(2010)\citenamefont{Zhao, M{\"{u}}ller,
  Hammerer, and Zoller}}]{PhysRevA.81.052329}
\bibinfo{author}{\bibfnamefont{B.}~\bibnamefont{Zhao}},
  \bibinfo{author}{\bibfnamefont{M.}~\bibnamefont{M{\"{u}}ller}},
  \bibinfo{author}{\bibfnamefont{K.}~\bibnamefont{Hammerer}}, \bibnamefont{and}
  \bibinfo{author}{\bibfnamefont{P.}~\bibnamefont{Zoller}},
  \bibinfo{journal}{Phys. Rev. A} \textbf{\bibinfo{volume}{81}},
  \bibinfo{pages}{052329} (\bibinfo{year}{2010}).

\bibitem[{\citenamefont{Brion et~al.}(2012)\citenamefont{Brion, Carlier,
  Akulin, and M$\backslash$olmer}}]{PhysRevA.85.042324}
\bibinfo{author}{\bibfnamefont{E.}~\bibnamefont{Brion}},
  \bibinfo{author}{\bibfnamefont{F.}~\bibnamefont{Carlier}},
  \bibinfo{author}{\bibfnamefont{V.~M.} \bibnamefont{Akulin}},
  \bibnamefont{and}
  \bibinfo{author}{\bibfnamefont{K.}~\bibnamefont{M{\o}lmer}},
  \bibinfo{journal}{Phys. Rev. A} \textbf{\bibinfo{volume}{85}},
  \bibinfo{pages}{042324} (\bibinfo{year}{2012}).

\bibitem[{\citenamefont{Solmeyer et~al.}(2015)\citenamefont{Solmeyer, Li, and
  Quraishi}}]{Solmeyer2015a}
\bibinfo{author}{\bibfnamefont{N.}~\bibnamefont{Solmeyer}},
  \bibinfo{author}{\bibfnamefont{X.}~\bibnamefont{Li}}, \bibnamefont{and}
  \bibinfo{author}{\bibfnamefont{Q.}~\bibnamefont{Quraishi}}, 
\bibinfo{journal}{Phys. Rev. A} \textbf{\bibinfo{volume}{93}},
  \bibinfo{pages}{042301} (\bibinfo{year}{2016}).

\bibitem[{\citenamefont{Jaksch et~al.}(2000)\citenamefont{Jaksch, Cirac,
  Zoller, Rolston, C{\^{o}}t{\'{e}}, and Lukin}}]{PhysRevLett.85.2208}
\bibinfo{author}{\bibfnamefont{D.}~\bibnamefont{Jaksch}},
  \bibinfo{author}{\bibfnamefont{J.~I.} \bibnamefont{Cirac}},
  \bibinfo{author}{\bibfnamefont{P.}~\bibnamefont{Zoller}},
  \bibinfo{author}{\bibfnamefont{S.~L.} \bibnamefont{Rolston}},
  \bibinfo{author}{\bibfnamefont{R.}~\bibnamefont{C{\^{o}}t{\'{e}}}},
  \bibnamefont{and} \bibinfo{author}{\bibfnamefont{M.~D.} \bibnamefont{Lukin}},
  \bibinfo{journal}{Phys. Rev. Lett.} \textbf{\bibinfo{volume}{85}},
  \bibinfo{pages}{2208} (\bibinfo{year}{2000}).

\bibitem[{\citenamefont{Lukin et~al.}(2001)\citenamefont{Lukin, Fleischhauer,
  Cote, Duan, Jaksch, Cirac, and Zoller}}]{PhysRevLett.87.037901}
\bibinfo{author}{\bibfnamefont{M.~D.} \bibnamefont{Lukin}},
  \bibinfo{author}{\bibfnamefont{M.}~\bibnamefont{Fleischhauer}},
  \bibinfo{author}{\bibfnamefont{R.}~\bibnamefont{Cote}},
  \bibinfo{author}{\bibfnamefont{L.~M.} \bibnamefont{Duan}},
  \bibinfo{author}{\bibfnamefont{D.}~\bibnamefont{Jaksch}},
  \bibinfo{author}{\bibfnamefont{J.~I.} \bibnamefont{Cirac}}, \bibnamefont{and}
  \bibinfo{author}{\bibfnamefont{P.}~\bibnamefont{Zoller}},
  \bibinfo{journal}{Phys. Rev. Lett.} \textbf{\bibinfo{volume}{87}},
  \bibinfo{pages}{037901} (\bibinfo{year}{2001}).

\bibitem[{\citenamefont{Isenhower et~al.}(2010)\citenamefont{Isenhower, Urban,
  Zhang, Gill, Henage, Johnson, Walker, and Saffman}}]{Isenhower_PRL2010}
\bibinfo{author}{\bibfnamefont{L.}~\bibnamefont{Isenhower}},
  \bibinfo{author}{\bibfnamefont{E.}~\bibnamefont{Urban}},
  \bibinfo{author}{\bibfnamefont{X.~L.} \bibnamefont{Zhang}},
  \bibinfo{author}{\bibfnamefont{A.~T.} \bibnamefont{Gill}},
  \bibinfo{author}{\bibfnamefont{T.}~\bibnamefont{Henage}},
  \bibinfo{author}{\bibfnamefont{T.~A.} \bibnamefont{Johnson}},
  \bibinfo{author}{\bibfnamefont{T.~G.} \bibnamefont{Walker}},
  \bibnamefont{and} \bibinfo{author}{\bibfnamefont{M.}~\bibnamefont{Saffman}},
  \bibinfo{journal}{Phys. Rev. Lett.} \textbf{\bibinfo{volume}{104}},
  \bibinfo{pages}{010503} (\bibinfo{year}{2010}).

\bibitem[{\citenamefont{Zhang et~al.}(2012)\citenamefont{Zhang, Gill,
  Isenhower, Walker, and Saffman}}]{Zhang_PRA12_FideRydGate}
\bibinfo{author}{\bibfnamefont{X.~L.} \bibnamefont{Zhang}},
  \bibinfo{author}{\bibfnamefont{A.~T.} \bibnamefont{Gill}},
  \bibinfo{author}{\bibfnamefont{L.}~\bibnamefont{Isenhower}},
  \bibinfo{author}{\bibfnamefont{T.~G.} \bibnamefont{Walker}},
  \bibnamefont{and} \bibinfo{author}{\bibfnamefont{M.}~\bibnamefont{Saffman}},
  \bibinfo{journal}{Phys. Rev. A} \textbf{\bibinfo{volume}{85}},
  \bibinfo{pages}{042310} (\bibinfo{year}{2012}).

\bibitem[{\citenamefont{M{\"{u}}ller et~al.}(2014)\citenamefont{M{\"{u}}ller,
  Murphy, Montangero, Calarco, Grangier, and Browaeys}}]{Muller2014}
\bibinfo{author}{\bibfnamefont{M.~M.} \bibnamefont{M{\"{u}}ller}},
  \bibinfo{author}{\bibfnamefont{M.}~\bibnamefont{Murphy}},
  \bibinfo{author}{\bibfnamefont{S.}~\bibnamefont{Montangero}},
  \bibinfo{author}{\bibfnamefont{T.}~\bibnamefont{Calarco}},
  \bibinfo{author}{\bibfnamefont{P.}~\bibnamefont{Grangier}}, \bibnamefont{and}
  \bibinfo{author}{\bibfnamefont{A.}~\bibnamefont{Browaeys}},
  \bibinfo{journal}{Phys. Rev. A} \textbf{\bibinfo{volume}{89}},
  \bibinfo{pages}{032334} (\bibinfo{year}{2014}).

\bibitem[{\citenamefont{Maller et~al.}(2015)\citenamefont{Maller, Lichtman,
  Xia, Sun, Piotrowicz, Carr, Isenhower, and Saffman}}]{Maller2015}
\bibinfo{author}{\bibfnamefont{K.~M.} \bibnamefont{Maller}},
  \bibinfo{author}{\bibfnamefont{M.~T.} \bibnamefont{Lichtman}},
  \bibinfo{author}{\bibfnamefont{T.}~\bibnamefont{Xia}},
  \bibinfo{author}{\bibfnamefont{Y.}~\bibnamefont{Sun}},
  \bibinfo{author}{\bibfnamefont{M.~J.} \bibnamefont{Piotrowicz}},
  \bibinfo{author}{\bibfnamefont{A.~W.} \bibnamefont{Carr}},
  \bibinfo{author}{\bibfnamefont{L.}~\bibnamefont{Isenhower}},
  \bibnamefont{and} \bibinfo{author}{\bibfnamefont{M.}~\bibnamefont{Saffman}},
  \bibinfo{journal}{Phys. Rev. A} \textbf{\bibinfo{volume}{92}},
  \bibinfo{pages}{022336} (\bibinfo{year}{2015}).

\bibitem[{\citenamefont{Su et~al.}(2017)\citenamefont{Su, Gao, Liang, and
  Zhang}}]{Su2017}
\bibinfo{author}{\bibfnamefont{S.-L.} \bibnamefont{Su}},
  \bibinfo{author}{\bibfnamefont{Y.}~\bibnamefont{Gao}},
  \bibinfo{author}{\bibfnamefont{E.}~\bibnamefont{Liang}}, \bibnamefont{and}
  \bibinfo{author}{\bibfnamefont{S.}~\bibnamefont{Zhang}},
  \bibinfo{journal}{Phys. Rev. A} \textbf{\bibinfo{volume}{95}},
  \bibinfo{pages}{022319} (\bibinfo{year}{2017}).

\bibitem[{\citenamefont{Petrosyan and M{\o}lmer}(2014)}]{Petrosyan2014}
\bibinfo{author}{\bibfnamefont{D.}~\bibnamefont{Petrosyan}} \bibnamefont{and}
  \bibinfo{author}{\bibfnamefont{K.}~\bibnamefont{M{\o}lmer}},
  \bibinfo{journal}{Phys. Rev. Lett.} \textbf{\bibinfo{volume}{113}},
  \bibinfo{pages}{123003} (\bibinfo{year}{2014}).

\bibitem[{\citenamefont{Barredo et~al.}(2014)\citenamefont{Barredo, Ravets,
  Labuhn, B{\'{e}}guin, Vernier, Nogrette, Lahaye, and Browaeys}}]{Barredo2014}
\bibinfo{author}{\bibfnamefont{D.}~\bibnamefont{Barredo}},
  \bibinfo{author}{\bibfnamefont{S.}~\bibnamefont{Ravets}},
  \bibinfo{author}{\bibfnamefont{H.}~\bibnamefont{Labuhn}},
  \bibinfo{author}{\bibfnamefont{L.}~\bibnamefont{B{\'{e}}guin}},
  \bibinfo{author}{\bibfnamefont{A.}~\bibnamefont{Vernier}},
  \bibinfo{author}{\bibfnamefont{F.}~\bibnamefont{Nogrette}},
  \bibinfo{author}{\bibfnamefont{T.}~\bibnamefont{Lahaye}}, \bibnamefont{and}
  \bibinfo{author}{\bibfnamefont{A.}~\bibnamefont{Browaeys}},
  \bibinfo{journal}{Phys. Rev. Lett.} \textbf{\bibinfo{volume}{112}},
  \bibinfo{pages}{183002} (\bibinfo{year}{2014}).

\bibitem[{\citenamefont{Zeng et~al.}(2017)\citenamefont{Zeng, Xu, He, Liu, Liu,
  Wang, Papoular, Shlyapnikov, and Zhan}}]{Zeng2017}
\bibinfo{author}{\bibfnamefont{Y.}~\bibnamefont{Zeng}},
  \bibinfo{author}{\bibfnamefont{P.}~\bibnamefont{Xu}},
  \bibinfo{author}{\bibfnamefont{X.~D.} \bibnamefont{He}},
  \bibinfo{author}{\bibfnamefont{Y.~Y.} \bibnamefont{Liu}},
  \bibinfo{author}{\bibfnamefont{M.}~\bibnamefont{Liu}},
  \bibinfo{author}{\bibfnamefont{J.}~\bibnamefont{Wang}},
  \bibinfo{author}{\bibfnamefont{D.~J.} \bibnamefont{Papoular}},
  \bibinfo{author}{\bibfnamefont{G.~V.} \bibnamefont{Shlyapnikov}},
  \bibnamefont{and} \bibinfo{author}{\bibfnamefont{M.~S.} \bibnamefont{Zhan}},
  \bibinfo{journal}{arXiv: 1702.00349}.

\bibitem[{\citenamefont{Vitanov et~al.}(2017)\citenamefont{Vitanov, Rangelov,
  Shore, and Bergmann}}]{Vitanov2017}
\bibinfo{author}{\bibfnamefont{N.~V.} \bibnamefont{Vitanov}},
  \bibinfo{author}{\bibfnamefont{A.~A.} \bibnamefont{Rangelov}},
  \bibinfo{author}{\bibfnamefont{B.~W.} \bibnamefont{Shore}}, \bibnamefont{and}
  \bibinfo{author}{\bibfnamefont{K.}~\bibnamefont{Bergmann}},
  \bibinfo{journal}{Rev. Mod. Phys.} \textbf{\bibinfo{volume}{89}},
  \bibinfo{pages}{015006} (\bibinfo{year}{2017}).

\bibitem[{\citenamefont{Beterov et~al.}(2011)\citenamefont{Beterov, Tretyakov,
  Entin, Yakshina, Ryabtsev, MacCormick, and Bergamini}}]{PhysRevA.84.023413}
\bibinfo{author}{\bibfnamefont{I.~I.} \bibnamefont{Beterov}},
  \bibinfo{author}{\bibfnamefont{D.~B.} \bibnamefont{Tretyakov}},
  \bibinfo{author}{\bibfnamefont{V.~M.} \bibnamefont{Entin}},
  \bibinfo{author}{\bibfnamefont{E.~A.} \bibnamefont{Yakshina}},
  \bibinfo{author}{\bibfnamefont{I.~I.} \bibnamefont{Ryabtsev}},
  \bibinfo{author}{\bibfnamefont{C.}~\bibnamefont{MacCormick}},
  \bibnamefont{and}
  \bibinfo{author}{\bibfnamefont{S.}~\bibnamefont{Bergamini}},
  \bibinfo{journal}{Phys. Rev. A} \textbf{\bibinfo{volume}{84}},
  \bibinfo{pages}{023413} (\bibinfo{year}{2011}).

\bibitem[{\citenamefont{Yan et~al.}(2011)\citenamefont{Yan, Cui, Zhang, and
  Wu}}]{Yan2011}
\bibinfo{author}{\bibfnamefont{D.}~\bibnamefont{Yan}},
  \bibinfo{author}{\bibfnamefont{C.~L.} \bibnamefont{Cui}},
  \bibinfo{author}{\bibfnamefont{M.}~\bibnamefont{Zhang}}, \bibnamefont{and}
  \bibinfo{author}{\bibfnamefont{J.~H.} \bibnamefont{Wu}},
  \bibinfo{journal}{Phys. Rev. A}
  \textbf{\bibinfo{volume}{84}}, \bibinfo{pages}{043405} (\bibinfo{year}{2011}),
 .

\bibitem[{\citenamefont{Qian et~al.}(2015)\citenamefont{Qian, Zhai, Zhang, and
  Zhang}}]{Qian2015a}
\bibinfo{author}{\bibfnamefont{J.}~\bibnamefont{Qian}},
  \bibinfo{author}{\bibfnamefont{J.}~\bibnamefont{Zhai}},
  \bibinfo{author}{\bibfnamefont{L.}~\bibnamefont{Zhang}}, \bibnamefont{and}
  \bibinfo{author}{\bibfnamefont{W.}~\bibnamefont{Zhang}},
  \bibinfo{journal}{Phys. Rev. A} \textbf{\bibinfo{volume}{91}},
  \bibinfo{pages}{013411} (\bibinfo{year}{2015}).

\bibitem[{\citenamefont{Tian et~al.}(2015)\citenamefont{Tian, Liu, Cui, and
  Wu}}]{Tian2015}
\bibinfo{author}{\bibfnamefont{X.-D.} \bibnamefont{Tian}},
  \bibinfo{author}{\bibfnamefont{Y.-M.} \bibnamefont{Liu}},
  \bibinfo{author}{\bibfnamefont{C.-L.} \bibnamefont{Cui}}, \bibnamefont{and}
  \bibinfo{author}{\bibfnamefont{J.-H.} \bibnamefont{Wu}},
  \bibinfo{journal}{Phys. Rev. A} \textbf{\bibinfo{volume}{92}},
  \bibinfo{pages}{063411} (\bibinfo{year}{2015}).

\bibitem[{\citenamefont{Petrosyan et~al.}(2015)\citenamefont{Petrosyan, Rao,
  and M{\o}lmer}}]{Petrosyan2015d}
\bibinfo{author}{\bibfnamefont{D.}~\bibnamefont{Petrosyan}},
  \bibinfo{author}{\bibfnamefont{D.~D.~B.} \bibnamefont{Rao}},
  \bibnamefont{and}
  \bibinfo{author}{\bibfnamefont{K.}~\bibnamefont{M{\o}lmer}},
  \bibinfo{journal}{Phys. Rev. A}
  \textbf{\bibinfo{volume}{91}}, \bibinfo{pages}{043402} (\bibinfo{year}{2015}).

\bibitem[{\citenamefont{M{\o}ller et~al.}(2008)\citenamefont{M{\o}ller, Madsen,
  and M{\o}lmer}}]{Moller2008}
\bibinfo{author}{\bibfnamefont{D.}~\bibnamefont{M{\o}ller}},
  \bibinfo{author}{\bibfnamefont{L.}~\bibnamefont{Madsen}}, \bibnamefont{and}
  \bibinfo{author}{\bibfnamefont{K.}~\bibnamefont{M{\o}lmer}},
  \bibinfo{journal}{Phys. Rev. Lett.} \textbf{\bibinfo{volume}{100}},
  \bibinfo{pages}{170504} (\bibinfo{year}{2008}).

\bibitem[{\citenamefont{Goerz et~al.}(2014)\citenamefont{Goerz, Halperin,
  Aytac, Koch, and Whaley}}]{Goerz2014}
\bibinfo{author}{\bibfnamefont{M.~H.} \bibnamefont{Goerz}},
  \bibinfo{author}{\bibfnamefont{E.~J.} \bibnamefont{Halperin}},
  \bibinfo{author}{\bibfnamefont{J.~M.} \bibnamefont{Aytac}},
  \bibinfo{author}{\bibfnamefont{C.~P.} \bibnamefont{Koch}}, \bibnamefont{and}
  \bibinfo{author}{\bibfnamefont{K.~B.} \bibnamefont{Whaley}},
  \bibinfo{journal}{Phys. Rev. A} \textbf{\bibinfo{volume}{90}},
  \bibinfo{pages}{032329} (\bibinfo{year}{2014}).

\bibitem[{\citenamefont{Beterov et~al.}(2013)\citenamefont{Beterov, Saffman,
  Yakshina, Zhukov, Tretyakov, Entin, Ryabtsev, Mansell, MacCormick, Bergamini
  et~al.}}]{PhysRevA.88.010303}
\bibinfo{author}{\bibfnamefont{I.~I.} \bibnamefont{Beterov}},
  \bibinfo{author}{\bibfnamefont{M.}~\bibnamefont{Saffman}},
  \bibinfo{author}{\bibfnamefont{E.~A.} \bibnamefont{Yakshina}},
  \bibinfo{author}{\bibfnamefont{V.~P.} \bibnamefont{Zhukov}},
  \bibinfo{author}{\bibfnamefont{D.~B.} \bibnamefont{Tretyakov}},
  \bibinfo{author}{\bibfnamefont{V.~M.} \bibnamefont{Entin}},
  \bibinfo{author}{\bibfnamefont{I.~I.} \bibnamefont{Ryabtsev}},
  \bibinfo{author}{\bibfnamefont{C.~W.} \bibnamefont{Mansell}},
  \bibinfo{author}{\bibfnamefont{C.}~\bibnamefont{MacCormick}},
  \bibinfo{author}{\bibfnamefont{S.}~\bibnamefont{Bergamini}},
  \bibnamefont{et~al.}, \bibinfo{journal}{Phys. Rev. A}
  \textbf{\bibinfo{volume}{88}}, \bibinfo{pages}{010303}
  (\bibinfo{year}{2013}).

\bibitem[{\citenamefont{Zheng and Brun}(2012)}]{PhysRevA.86.032323}
\bibinfo{author}{\bibfnamefont{Y.-C.} \bibnamefont{Zheng}} \bibnamefont{and}
  \bibinfo{author}{\bibfnamefont{T.~A.} \bibnamefont{Brun}},
  \bibinfo{journal}{Phys. Rev. A} \textbf{\bibinfo{volume}{86}},
  \bibinfo{pages}{032323} (\bibinfo{year}{2012}).

\bibitem[{\citenamefont{Zheng}(2015)}]{Zheng2015}
\bibinfo{author}{\bibfnamefont{S.-B.} \bibnamefont{Zheng}},
  \bibinfo{journal}{Phys. Rev. A}
  \textbf{\bibinfo{volume}{91}}, \bibinfo{pages}{052117} (\bibinfo{year}{2015}).

\bibitem[{\citenamefont{Zheng et~al.}(2016)\citenamefont{Zheng, Yang, and
  Nori}}]{Zheng2016}
\bibinfo{author}{\bibfnamefont{S.-B.} \bibnamefont{Zheng}},
  \bibinfo{author}{\bibfnamefont{C.~P.} \bibnamefont{Yang}}, \bibnamefont{and}
  \bibinfo{author}{\bibfnamefont{F.}~\bibnamefont{Nori}},
  \bibinfo{journal}{Phys. Rev. A}
  \textbf{\bibinfo{volume}{93}}, \bibinfo{pages}{032313} (\bibinfo{year}{2016}).

\bibitem[{\citenamefont{Zheng}(2005)}]{Zheng_PRL2005}
\bibinfo{author}{\bibfnamefont{S.-B.} \bibnamefont{Zheng}},
  \bibinfo{journal}{Phys. Rev. Lett.} \textbf{\bibinfo{volume}{95}},
  \bibinfo{pages}{080502} (\bibinfo{year}{2005}).

\bibitem[{\citenamefont{Saffman and M{\o}lmer}(2008)}]{Saffman2008}
\bibinfo{author}{\bibfnamefont{M.}~\bibnamefont{Saffman}} \bibnamefont{and}
  \bibinfo{author}{\bibfnamefont{K.}~\bibnamefont{M{\o}lmer}},
  \bibinfo{journal}{Phys. Rev. A} \textbf{\bibinfo{volume}{78}},
  \bibinfo{pages}{012336} (\bibinfo{year}{2008}).

\bibitem[{\citenamefont{Saffman and
  M$\backslash$olmer}(2009)}]{PhysRevLett.102.240502}
\bibinfo{author}{\bibfnamefont{M.}~\bibnamefont{Saffman}} \bibnamefont{and}
  \bibinfo{author}{\bibfnamefont{K.}~\bibnamefont{M{\o}lmer}},
  \bibinfo{journal}{Phys. Rev. Lett.} \textbf{\bibinfo{volume}{102}},
  \bibinfo{pages}{240502} (\bibinfo{year}{2009}).

\bibitem[{\citenamefont{Hankin et~al.}(2014)\citenamefont{Hankin, Jau,
  Parazzoli, Chou, Armstrong, Landahl, and Biedermann}}]{Hankin2014b}
\bibinfo{author}{\bibfnamefont{A.~M.} \bibnamefont{Hankin}},
  \bibinfo{author}{\bibfnamefont{Y.-Y.} \bibnamefont{Jau}},
  \bibinfo{author}{\bibfnamefont{L.~P.} \bibnamefont{Parazzoli}},
  \bibinfo{author}{\bibfnamefont{C.~W.} \bibnamefont{Chou}},
  \bibinfo{author}{\bibfnamefont{D.~J.} \bibnamefont{Armstrong}},
  \bibinfo{author}{\bibfnamefont{A.~J.} \bibnamefont{Landahl}},
  \bibnamefont{and} \bibinfo{author}{\bibfnamefont{G.~W.}
  \bibnamefont{Biedermann}}, \bibinfo{journal}{Phys. Rev. A}
  \textbf{\bibinfo{volume}{89}}, \bibinfo{pages}{033416}
  (\bibinfo{year}{2014}).

\bibitem[{\citenamefont{Maxwell et~al.}(2013)\citenamefont{Maxwell, Szwer,
  Paredes-Barato, Busche, Pritchard, Gauguet, Weatherill, Jones, and
  Adams}}]{PhysRevLett.110.103001}
\bibinfo{author}{\bibfnamefont{D.}~\bibnamefont{Maxwell}},
  \bibinfo{author}{\bibfnamefont{D.~J.} \bibnamefont{Szwer}},
  \bibinfo{author}{\bibfnamefont{D.}~\bibnamefont{Paredes-Barato}},
  \bibinfo{author}{\bibfnamefont{H.}~\bibnamefont{Busche}},
  \bibinfo{author}{\bibfnamefont{J.~D.} \bibnamefont{Pritchard}},
  \bibinfo{author}{\bibfnamefont{A.}~\bibnamefont{Gauguet}},
  \bibinfo{author}{\bibfnamefont{K.~J.} \bibnamefont{Weatherill}},
  \bibinfo{author}{\bibfnamefont{M.~P.~A.} \bibnamefont{Jones}},
  \bibnamefont{and} \bibinfo{author}{\bibfnamefont{C.~S.} \bibnamefont{Adams}},
  \bibinfo{journal}{Phys. Rev. Lett.} \textbf{\bibinfo{volume}{110}},
  \bibinfo{pages}{103001} (\bibinfo{year}{2013}).

\bibitem[{\citenamefont{Barredo et~al.}(2015)\citenamefont{Barredo, Labuhn,
  Ravets, Lahaye, Browaeys, and Adams}}]{Barredo2015}
\bibinfo{author}{\bibfnamefont{D.}~\bibnamefont{Barredo}},
  \bibinfo{author}{\bibfnamefont{H.}~\bibnamefont{Labuhn}},
  \bibinfo{author}{\bibfnamefont{S.}~\bibnamefont{Ravets}},
  \bibinfo{author}{\bibfnamefont{T.}~\bibnamefont{Lahaye}},
  \bibinfo{author}{\bibfnamefont{A.}~\bibnamefont{Browaeys}}, \bibnamefont{and}
  \bibinfo{author}{\bibfnamefont{C.~S.} \bibnamefont{Adams}},
  \bibinfo{journal}{Phys. Rev. Lett.} \textbf{\bibinfo{volume}{114}},
  \bibinfo{pages}{113002} (\bibinfo{year}{2015}).

\end{thebibliography}

\end{document}